\documentclass{ws-mpla}

\usepackage{graphicx, epsfig}
\usepackage{color}
\usepackage{amsmath}
\usepackage{hyperref}
\newcommand{\be}{\begin{equation}}
\newcommand{\ee}{\end{equation}}
\newcommand{\bea}{\begin{eqnarray}}
\newcommand{\eea}{\end{eqnarray}}

\begin{document}

\markboth{Dejan Stojkovic}
{Vanishing dimensions: Review}

\catchline{}{}{}{}{}

\title{Vanishing dimensions: Review}

\author{\footnotesize Dejan Stojkovic\footnote{
HEPCOS, Department of Physics, SUNY at Buffalo, Buffalo, NY 14260-1500}}

\address{HEPCOS, Department of Physics, SUNY at Buffalo, Buffalo, NY 14260-1500\\
and  Perimeter Institute for Theoretical Physics, 31 Caroline St. N., Waterloo, ON, N2L 2Y5, Canada \\
ds77@buffalo.edu}

\maketitle

\pub{Received (Day Month Year)}{Revised (Day Month Year)}

\begin{abstract}
We review a growing theoretical motivation and evidence that the number of dimensions actually reduces at high energies. This reduction can happen near the Planck scale, or much before, the dimensions that are reduced can be effective, spectral, topological or the usual dimensions, but many things points toward the fact that the high energy theories appear to propagate in a lower dimensional space, rather than a higher dimensional one. We will concentrate on a particular scenario of ``vanishing" or ``evolving dimensions" where the dimensions open up as we increase the length scale that we are probing, but will also mention related models that point to the same direction, i.e. the causal dynamical triangulation, asymptotic safety, as well as evidence coming from a non-commutative quantum theories, the Wheeler-DeWitt equation and phenomenon of ``asymptotic silence". It is intriguing that experimental evidence for the high energy dimensional reduction may already exists - a statistically significant planar alignment of events with energies higher than TeV has been observed in high altitude cosmic ray experiments. A convincing evidence for dimensional reduction may be found in future in collider experiments and gravity waves observatories.
\end{abstract}

\ccode{PACS Nos.: include PACS Nos.}

\section{Introduction}

The idea that the number of dimensions in our universe is different from $3+1$ is not new. In order to unite gravity with electromagnetism, Kaluza and Klein proposed a $4+1$-dimensional universe. To unite all of the fundamental interactions, Kaluza-Klein models with up to $11$ spacetime dimensions were proposed. In string theory, self-consistency requires that the total number of dimensions is $26$ or $10$. The underlying idea is that our world is fundamentally higher dimensional, while it only appears to be $3+1$-dimensional at low energies, when we look at it with low resolution observations and experiments, which are unable to probe the short-distance physics which is higher-dimensional. It is somewhat surprising that very little effort was invested in the opposite direction, i.e. in the idea that our universe is fundamentally lower dimensional at high energies, while our $3+1$-dimensional world is only and effective low-energy picture. It is surprising since the advantage of having less dimensions at high energies where we encounter stubborn problems is obvious.

We believe that (with some exceptions) we understand our universe on scales approximately between $\sim 10^{-17}$ and $10^{27}$ cm's. The first scale corresponds to the energy scale of TeV which is the energy probed in the highest energy accelerators available so far. The second scale corresponds to the distance characteristic for super-clusters of galaxies, i.e. the scale at which cosmology kicks in. At scales shorter than $10^{-17}$ cm and larger than about a Gpc  $\sim 10^{27}$ cm, we are running into problems.

There exists a strong motivation to reduce the dimensionality of the spacetime at high energies (short distances). One of the most acute problems - the Standard Model hierarchy problem does not exists in $1+1$-dimensional spacetime since the corrections to the Higgs mass are only logarithmically divergent. There is no need for new particles and elaborate cancelation schemes. The coupling constant in QCD in $1+1$ and $2+1$ dimensions has positive dimension, and the theory is therefore super-renormalizable, i.e. only a finite set
of graphs need overall counter terms. Even the most elusive concept in modern physics - quantum gravity - is much more within the reach in lower dimensions. If the fundamental short scale physics is lower dimensional, there is no need to quantize $3+1$ dimensional gravity. Instead we should quantize  $2+1$ and $1+1$ dimensional gravity, which are, by comparison, much easier tasks.  $2+1$-dimensional general relativity has no local gravitational degrees of freedom, i.e. no gravitational waves in classical theory and no gravitons in quantum theory. Gravity is then completely determined by the local distribution of masses. The number of degrees of freedom in such a theory is finite, quantum field theory reduces to quantum mechanics and the problem of non-renormalizability disappears \cite{Carlip:1995zj}. For the reason of simplicity, $1+1$ dimensional gravity is even more attractive. Einstein's action in $1+1$  dimensional spacetime is a topological constant (Euler's characteristic of the manifold in question) and the theory is trivial (unless augmented by some additional fields). Models of gravity in $1+1$ dimensions are completely solvable \cite{Klosch:1997md,LouisMartinez:1993cc} and considerable work has been done on their quantum aspects \cite{Grumiller:2006ja,Zaslavskii:2003eu,Giddings:1992ae,Callan:1992rs,Bogojevic:1998ma}.

On intermediate scales between $10^{-17}$ cm and a Gpc, we know pretty well that our space is three dimensional. However, there is some  motivation to change the dimensionality of the spacetime on larger scales, comparable to the present cosmological horizon. Such ideas have been explored in a class of  brane-world models, known as {\it cascading gravity} \cite{cdgp,Hao:2014tsa}. An explicit construction that address the cosmological constant problem from a completely new perspective was introduced in \cite{Anchordoqui:2010er}. If the forth spatial dimension opens up at the current horizon scale, then an effective cosmological constant of the correct magnitude is induced without putting it into the equations by hand.

\begin{figure}
\centering
\includegraphics[width=4cm]{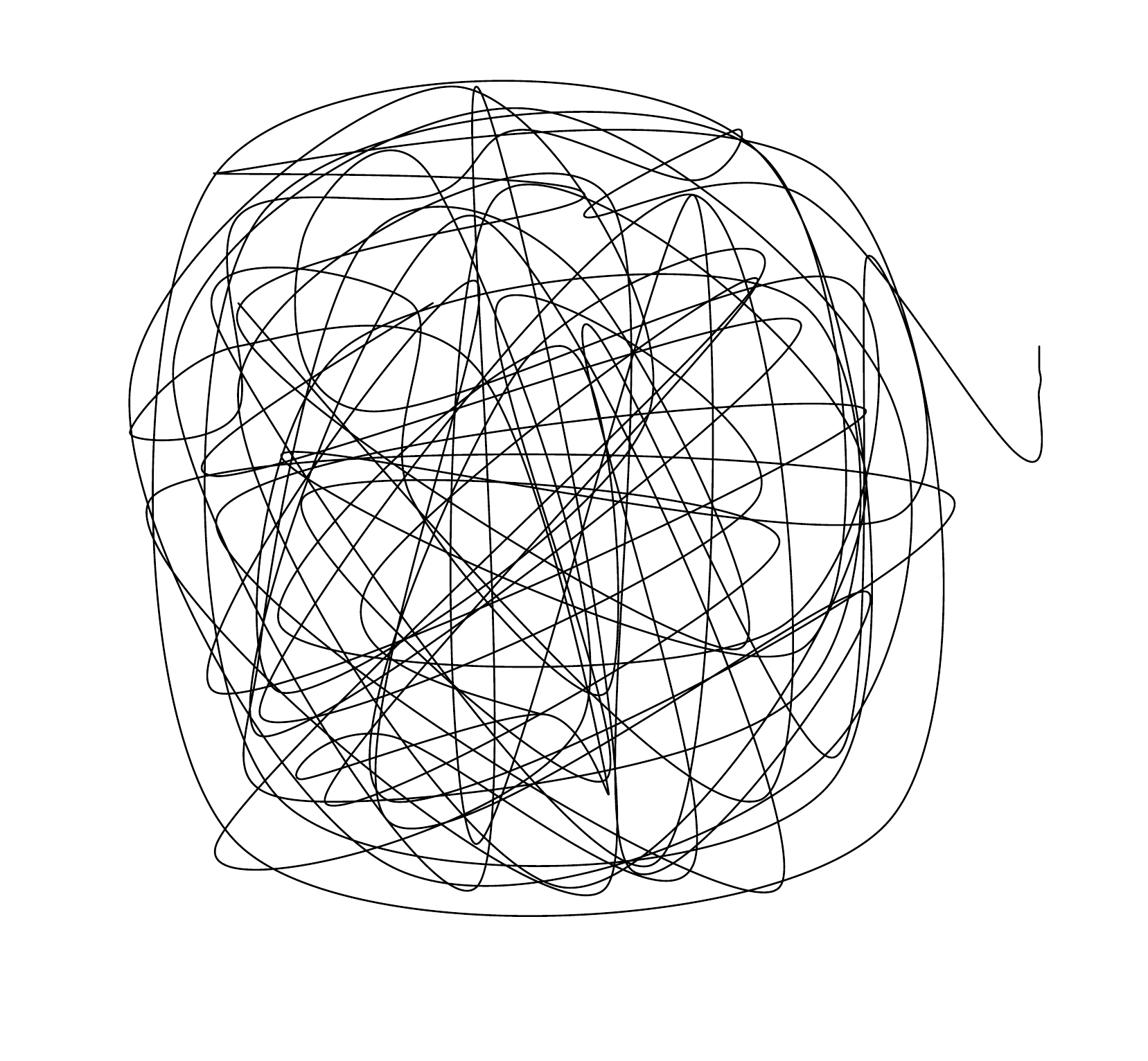}
\caption{An example of a structure which is one-dimensional on short scales while it appears effectively two-dimensional at large scales. It is sometimes easier to work with an ordered structure as in Fig.~\ref{lattice}.}
\label{1d2d} \end{figure}

\begin{figure}[h]
\center{\scalebox{0.25}{\includegraphics{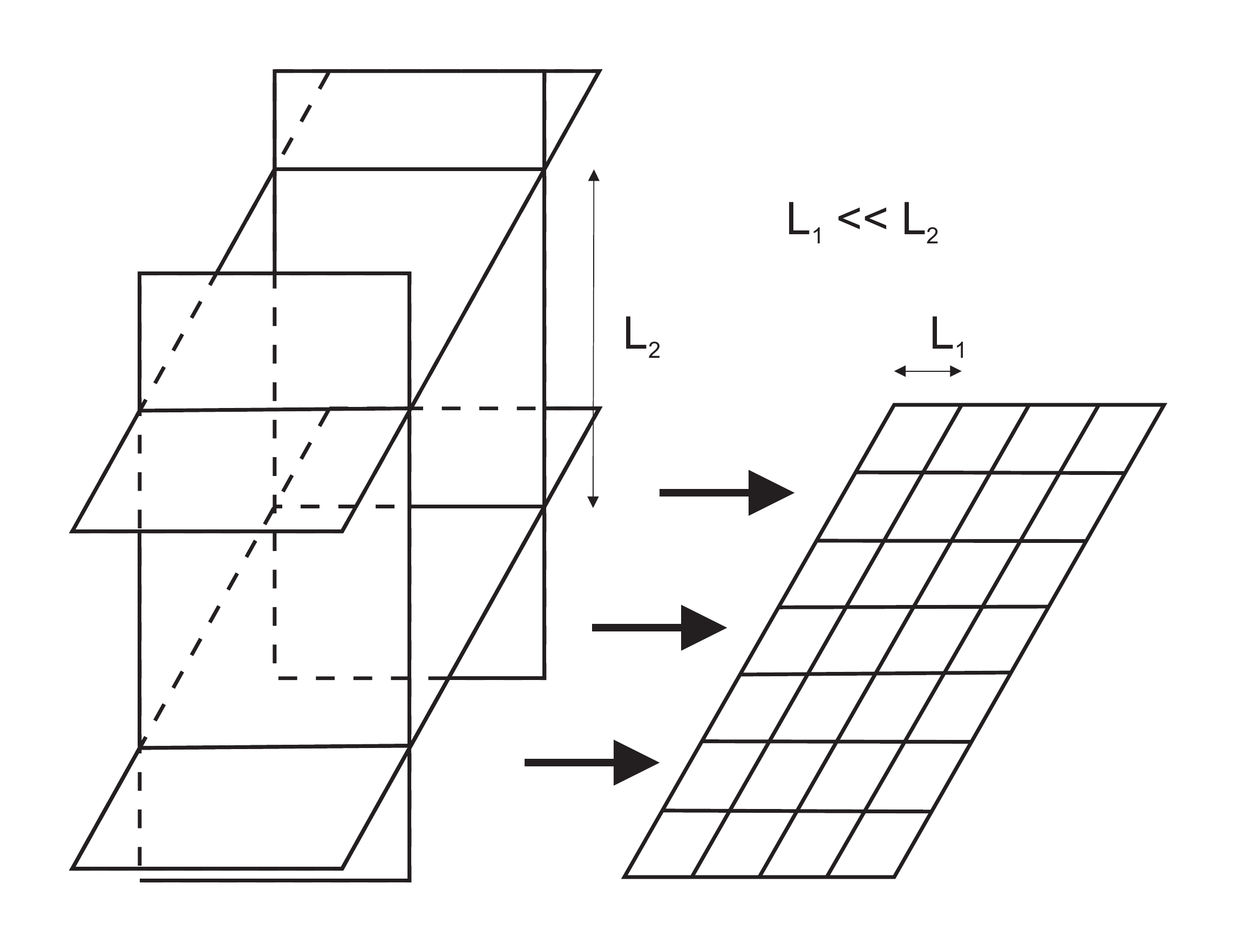}}}
\caption{A spacetime with an ordered lattice structure. Space structure is 1D on scales
  of the order of $L_1$, while it appears
  effectively 2D on scales much larger than $L_1$ but
   shorter than $L_2$. At scales much larger than
  $L_2$, the structure appears effectively 3D. Following this hierarchy, at even larger scales, say $L_3$, yet another dimension opens up and the structure appears 4D (not shown in the picture).}
\label{lattice}
\end{figure}

Once we notice that changing the dimensionality of our space at shortest and largest distances has manifold advantages, the most natural model with desired properties would be the one in which the effective number of dimensions increases with the length scale. At the shortest distances at which our space appears as continuum, the space is one-dimensional. At a certain critical length scale, the space becomes effectively two-dimensional (see Fig.~\ref{1d2d}). At the scale of about $10^{-17}$ cm, the space becomes effectively three-dimensional. Finally, at the scales of about a Gpc, the space becomes effectively four-dimensional. In principle, this hierarchy does not need to stop {\it a priori} at any finite number. However, it would be interesting for other reasons if this construct stops at $10$ or $26$ effective dimensions. In a dynamical picture where the universe starts from zero size and then grows, there is no background with the fixed number of dimensions in which the universe expands. The expanding universe encounters different number of dimensions during its evolution \cite{Stojkovic:2013lga}. For the sake of simplicity, it is always easier to work with a regular ordered lattice (Fig. \ref{lattice})  than with a random structure. Particles with the large momenta (short wavelengths) can probe the lower dimensional structure of the lattice. For example, particles whose wavelength is of the order of $L_1$ in Fig. (\ref{lattice}) move along the one-dimensional line. Particles whose wavelength is much longer than $L_1$ but shorter than $L_2$ move effectively in a two-dimensional space. In everyday three-dimensional life, we experience wavelengths much longer than $L_2$ but  much shorter than $L_3$.  Following the same hierarchy, the largest structures in our universe (of the size comparable to our horizon) may effectively be higher dimensional.

In this review, we will discuss the advantages of the dimensional reduction at high energies, point out that possible experimental evidence coming from high altitude cosmic ray experiments may already exist \cite{et1986,Mukhamedshin:2005nr,Antoni:2005ce}, and discuss predictions for the collider experiments and gravitational wave observatories. We will describe an ``ordered lattice" ad hoc model where particles of different energy see different number of dimensions, and also setup a Lagrangian for the concrete stringy model which captures the main idea of energy/temperature dependant number of dimensions. We will also point out that there is independent theoretical evidence which suggests that the short-distance spatial dimensionality is less than the macroscopically-observed three, namely "Causal dynamical triangulation", "Asymptotic safety" etc.
This review heavily relies on the previous publications by the author and his collaborators, and includes some work in progress.

\section{Removing ultraviolet divergences}
There exist a strong motivation for changing the dimensionality of the spacetime at small distances. One of the most acute problems connected with ultraviolet divergences is the Standard Model hierarchy problem.  The Higgs Lagrangian together with Yukawa couplings in the Standard Model is
\begin{equation}
L_H=D_\nu \Phi ^\dagger D^\nu \Phi - \mu^2 \Phi^\dagger \Phi + \frac{\lambda}{2}(\Phi^\dagger \Phi)^2 - \sum_f g_f \Phi \bar{\psi_f} \psi_f
\end{equation}
where $\Phi$ is the Higgs field, $g_f^2=m_f^2 /v^2$,  $\lambda = m_H^2/(2v^2)$, $m_f$ and $m_H$ are the fermion and Higgs mass respectively, while $v$ is the Higgs field vacuum expectation value.
If we consider all 1 loop one-particle-irreducible diagrams, then we find that the Higgs self energy comes from the three types of diagrams. In $3+1$ dimensions, all of these terms are quadratically divergent with the cut-off scale $\Lambda$ at which new physics appears. The contribution of fermions, gauge bosons and the Higgs itself in the loop are respectively
\begin{eqnarray} \label{terms}
&&  i \frac{g_f^2}{2}\int^\Lambda \frac{d^4k}{(2\pi)^4} tr(\frac{i}{\not k -m_f}\frac{i}{\not k+\not{p} -m_f})   \sim -\Lambda^2 \, \frac{g_f^2}{32\pi^2}  \nonumber \\
&&  \ \ \ \ \ \ \ i \frac{g^2}{4} \int^\Lambda \frac{d^4k}{(2\pi)^4} \frac{1}{k^2-m_g^2} \ \sim \ \Lambda^2 \, \frac{g^2}{64\pi^2} \\
&& \ \ \ \ \ \ \
i 3\lambda \int^\Lambda \frac{d^4k}{(2\pi)^4} \frac{1}{k^2-m_H^2} \ \sim \ \Lambda^2 \, \frac{3\lambda}{16\pi^2} \nonumber
\end{eqnarray}
Here, $g$ is the gauge coupling constant, $g^2=2m_g^2/v^2$, while
$m_g$ is the mass of gauge boson.
The quadratic divergence implies that a very special cancelation needs to happen between the bare Higgs mass and the corrections, unless the scale of new physics $\Lambda$ is very close to the electroweak scale.
\begin{figure}
\center{\includegraphics[width=1.5in]{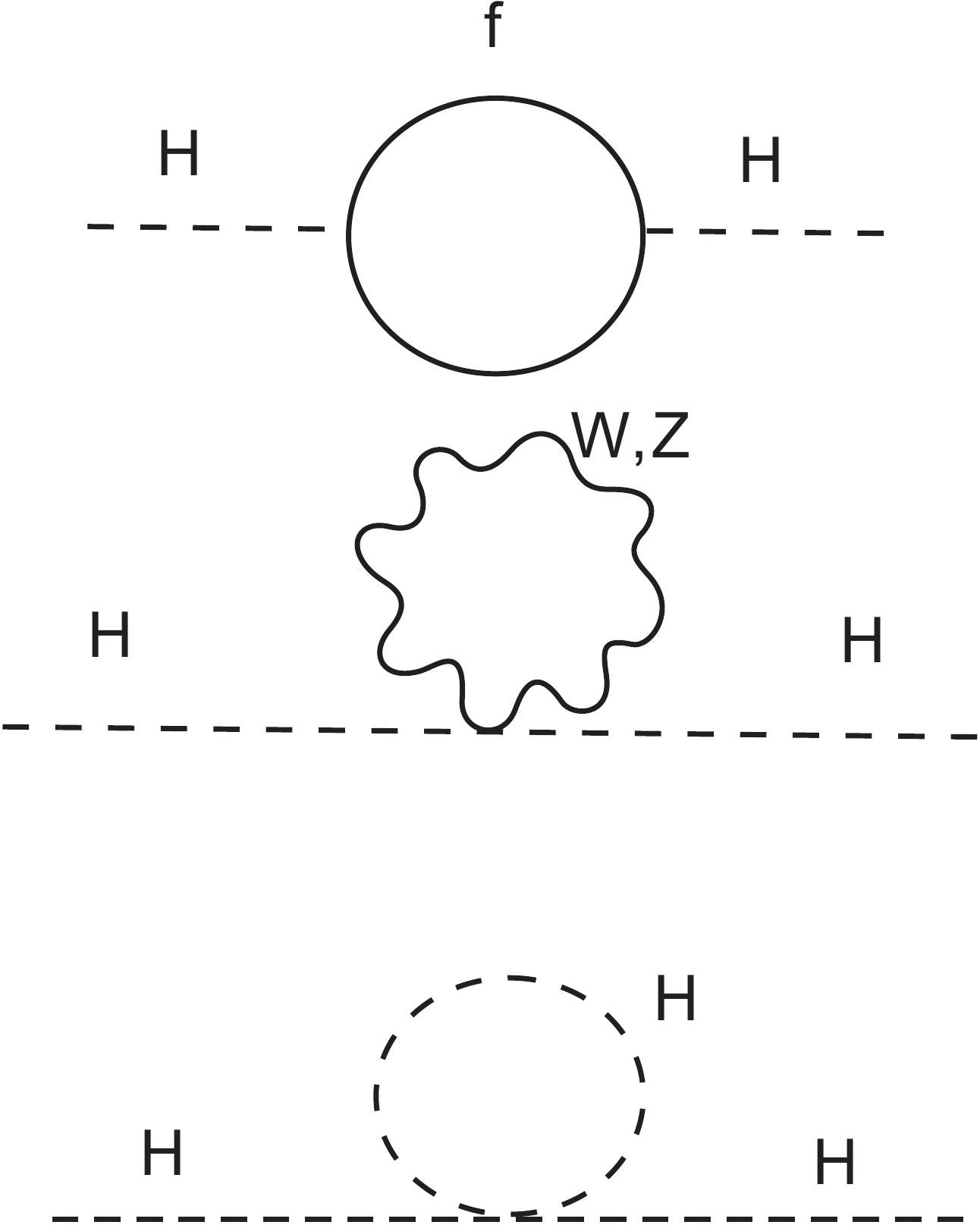}}
 \caption{Loop corrections to the Higgs mass: fermion, gauge boson and Higgs loop. They are quadratically divergent in $3+1$ dimensions, linearly in $2+1$ and only logarithmically in $1+1$ dimension.}
\label{1-loop}
\end{figure}

Solutions to the hierarchy problem proposed so far postulate new physics not very far from the electroweak scale.
An alternative approach would be to keep the Standard Model physics, and change the dimensionality of the background on which the model is defined. For example, in $2+1$ dimensional spacetime all of the terms in (\ref{terms}) are only linearly divergent

\begin{eqnarray}
&& i \frac{g_f^2}{2}\int^\Lambda \frac{d^3k}{(2\pi)^3} tr(\frac{i}{\not k -m_f}\frac{i}{\not k+\not{p} -m_f}) \ \sim \ -\Lambda \, \frac{g_f^2}{4\pi^2}  \nonumber \\
&&  \frac{g^2}{4} \int^\Lambda \frac{d^3k}{(2\pi)^3} \frac{1}{k^2-m_g^2} \ \sim \  \Lambda \, \frac{g^2}{8\pi^2}  \\
&&  i 3\lambda \int^\Lambda \frac{d^3k}{(2\pi)^3} \frac{1}{k^2-m_H^2} \ \sim \  \Lambda \, \frac{3\lambda}{2\pi^2}  \nonumber
\end{eqnarray}

Going further, in $1+1$ dimensional spacetime all of these terms are only logarithmically divergent

\begin{eqnarray}
&&  i \frac{g_f^2}{2}\int^\Lambda \frac{d^2k}{(2\pi)^2} tr(\frac{i}{\not k -m_f}\frac{i}{\not k+\not{p} -m_f}) \ \sim \ -\log(\Lambda/m_f) \, \frac{g_f^2}{4\pi}   \nonumber\\
&&  \frac{g^2}{4} \int^\Lambda \frac{d^2k}{(2\pi)^2} \frac{1}{k^2-m_g^2} \ \sim \ \log(\Lambda/m_g) \, \frac{g^2}{8\pi} \\
&&  i 3\lambda \int^\Lambda \frac{d^2k}{(2\pi)^2} \frac{1}{k^2-m_H^2} \ \sim \ \log(\Lambda/m_H) \, \frac{3\lambda}{2\pi}  \nonumber
\end{eqnarray}

Thus, keeping the Standard Model Lagrangian and lowering the dimensionality of the spacetime greatly improves the fine tuning problem in the Standard Model. In fact the dimensional regularization procedure tells us that the ultraviolet divergences in field theory are poles in the dimension plane. Lowering the dimensionality of the spacetime universally cures ultraviolet divergences in practically all of the field theories.

Possible implications of having less dimensions at higher energies are very important for the LHC physics.  There are three immediate and spectacular consequences of this model at the LHC, which should be observable if the dimensional crossover scale is  $\sim 1$~TeV, {\it i.e.\/}, within the reach of the machine: {\it (i)\/} cross section of hard scattering processes changes compared to that in the SM as the $Q^2$ becomes comparable with the crossover scale; {\it (ii)\/} $2 \to 4$ and higher order scattering processes at high energies become planar, resulting, {\it e.g.\/}, in four-jet events, where all jets are produced in one plane in their center-of-mass frame, thus strikingly different from standard QCD multijet events; {\it (iii)\/} under certain conditions, jets of sufficiently high energy may become elliptic in shape (for details see \cite{Anchordoqui:2010er,Anchordoqui:2010hi}).

\section{Gravity}

A satisfactory and self-consistent theory of quantum gravity remains one of the most difficult problems in modern physics.   Gravity in $3+1$ dimensions is complicated, highly nonlinear, perturbatively non-renormalizable theory. All of the attempts to successfully quantize gravity in $3+1$ dimensions met serious if not unsurmountable difficulties. However, quantum gravity is much more within the reach in lower dimensions. If the fundamental short scale physics is lower dimensional, and $3+1$ gravity is only an effective low energy theory, there is no need to quantize $3+1$ dimensional gravity. Instead we should quantize  $2+1$ and $1+1$ dimensional gravity. This is much easier task to accomplish. In any spacetime, the curvature tensor $R_{\mu \nu \rho \sigma }$ may be decomposed into a Ricci scalar $R$, Ricci tensor $R_{\mu \nu }$ and conformally invariant Weyl tensor $C_{\mu \nu \rho }^\sigma $. In $2+1$ dimensions the Weyl tensor vanishes and  $R_{\mu \nu \rho \sigma }$ can be expressed solely through $R_{\mu \nu }$ and $R$.  Explicitly
\be
R_{\mu \nu \rho \sigma } = \epsilon_{\mu \nu \alpha} \epsilon_{\rho \sigma \beta} G^{\alpha \beta}
\ee
This in turn implies that any solution of the vacuum Einstein's equations is locally flat. Thus, $2+1$  dimensional spacetime has no local gravitational degrees of freedom, i.e. no gravitational waves in classical theory and no gravitons in quantum theory. The number of degrees of freedom in such a theory is finite, quantum field theory reduces to quantum mechanics and the problem of non-renormalizability disappears \cite{Carlip:1995zj}. Obviously, $2+1$ dimensional gravity has much nicer structure than its $3+1$ dimensional cousin. For the reason of simplicity, $1+1$ dimensional gravity is even more attractive. Einstein's action in $1+1$  dimensional spacetime is a constant (Euler's characteristic of the manifold in question) and the theory is trivial (unless augmented by some additional fields). Models of gravity in $1+1$ dimensions are completely solvable \cite{Klosch:1997md,LouisMartinez:1993cc} and a lot of work has been done on their quantum aspects \cite{Grumiller:2006ja,Zaslavskii:2003eu,Giddings:1992ae,Callan:1992rs,Bogojevic:1998ma}.

\section{QCD}

If the spacetime is $2+1$ dimensional at distances shorter than $10^{-17}$ cm, then we expect some strong implications for the high energy scattering processes.
The structure of well established theories like Quantum Chromo Dynamics (QCD)  becomes much simpler in $2+1$ and $1+1$ dimensional spacetimes. The form of the QCD Lagrangian in $2+1$ dimensions is the same as in  $3+1$ dimensions
\begin{eqnarray}
&& {\cal L} = -\frac{1}{4}F^a_{\mu \nu } F^{a\mu \nu } + i \bar{\psi}\gamma^\mu (\partial_\mu + igA^a_\mu T^a)\psi  \\
&& F^a_{\mu \nu} = \partial_\mu A^a_\nu - \partial_\nu A^a_\mu + g f^{abc} A_\mu^b A_\nu ^b \, , \nonumber
\end{eqnarray}
except that the $\gamma$-matrices can be chosen to be proportional to two-dimensional Pauli matrices, while spinors $\psi$ are two-component spinors.
It is interesting that $2+1$ dimensional QCD is super-renormalizable, i.e. only a finite set of graphs need overall counter terms. This is a consequence of the fact that the coupling constant in this theory has positive dimension.
In  $2+1$ dimensional spacetime there is only one transverse dimension, so there is no arbitrarily high transverse angular momentum. This implies that there exist no Regge-like behavior due to  exchange of states of high spin which is characteristic in $3+1$ dimensional spacetime. For the LHC, it is certainly very important to calculate hadron-hadron scattering amplitude. It is not difficult to verify that the result is quite different from the standard one. The total cross section falls off like $1/\log s$, where $s$ is the center of mass energy squared. Characteristic Regge factor $s^\alpha, \alpha >0$ is completely absent \cite{Li:1994et}, in the strong contrast with $3+1$ dimensions.

For completeness, we mention $1+1$ dimensional QCD.
QCD in  $1+1$ dimensions is trivially asymptotically free, being super-renormalizable. The model on the infinite line has no gluon degrees of freedom, but it is a self-interacting fermion theory. In QCD on a circle, boundary conditions force the retention of quantum mechanical (zero-mode) gauge degrees of freedom in a Hamiltonian formulation. The dynamics of the zero modes lead to an elimination of fermionic non-singlet states from the spectrum in the continuum limit \cite{Engelhardt:1995qm}, thus practically eliminating color. While the details of $1+1$ and especially  $2+1$ dimensional QCD are very interesting for collider phenomenology, they  can not help (at least not directly) with the problem of quark confinement in a nucleon since the effective size of a nucleon (GeV$^{-1}$) is much larger than the critical distance at which the space appears $3+1$ dimensional (TeV$^{-1}$).

\section{Removing infrared divergences}

Infrared divergencies in field theories are usually eliminated by including contribution from infrared photons with infinite wavelength.
However, from the dimensional regularization point of view, infrared divergences, like their cousins ultraviolet ones, are also poles in the dimension plane. Therefore, changing the effective dimensionality of the spacetime will remove them trivially.
\begin{figure}
\center{\includegraphics[width=1in]{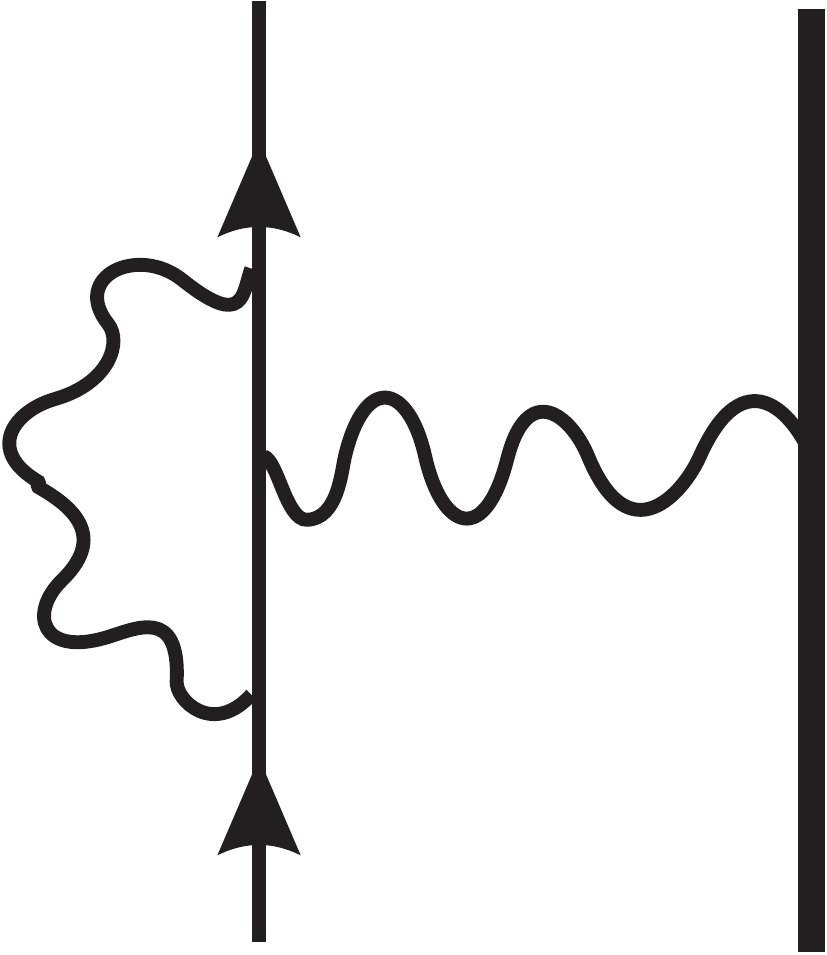}}
 \caption{The diagram of the infrared divergent electron-photon interaction.}
\label{infrared}
\end{figure}
For example, the electron-photon interaction cross section is divergent at low momenta (large wavelengths) of the virtual photon. The probability amplitude P for this process is
\begin{equation}
{\rm P} \approx \frac{\alpha}{\pi}\int^{|q|}_0 dk \frac{1}{k}I(v,v')
\end{equation}
Here, $k$ is the momentum of the virtual photon, $|q|$ is its  maximal value, while $v$ and $v'$ are initial and final velocity of the electron. $I(v,v')=dE/dk$ is the intensity, which is independent of $k$ when $k$ is small. The probability is obviously divergent at low $k$. We can remove this divergence by introducing an infrared momentum cutoff at very low energies. This can be ad-hock justified as a limitation of a finite size detector. Alternatively, we can include contribution from infrared photons with infinite wavelength, which will cancel the divergent terms. However, changing the effective dimensionality of the spacetime at large scales also removes this divergence. In our context, if the spacetime becomes effectively $4+1$ dimensional the integral becomes
\begin{equation}
{\rm P} \propto \int^{|q|}_0 dk
\end{equation}
This integral is not divergent as $k\rightarrow 0$ and the infrared divergence is removed. Thus, increasing the dimensionality of the spacetime effectively cures infrared divergences in a model independent way.

\section{Cosmology}

Changing the dimensionality of the spacetime at large distances may have some consequences for cosmology. This can happen if the length scale at which the fourth spatial dimension opens up is smaller than the present cosmological horizon. Available observational data indicate that our universe is going through a phase of accelerated expansion. To date it remains a mystery what is the driving force behind the acceleration. Data favor an equation of state of the cosmic fluid $p=-\rho$, corresponding to a constant energy density. The null hypothesis is that we are observing the action of the vacuum energy density (or cosmological constant). If it is indeed the cosmological constant we are seeing, it may represent the worst prediction ever made by a theory. Instead, the cosmological constant may just be a shadow that extra dimensions cast on our visible universe. To demonstrate this, we write down the metric of  $4+1$ dimensional  spacetime (with spatially isotropic $3$ dimensional slices) as
\be
ds^2 = e^{\nu }dt^2-e^{\omega }(dr^2+r^2d\Omega^2)-e^\mu d\psi^2 \, ,
\ee
where $\psi$ is the fourth spatial dimension, $d\Omega^2 \equiv d\theta^2+\sin^2\theta d\phi^2$, while  parameters $\nu, \mu$ and $\omega$ are arbitrary functions of $t$ and $\psi$. We can write down $4+1$ dimensional vacuum Einstein's equations
\be \label{GAB}
G_{AB}\equiv R_{AB}-\frac{1}{2}g_{AB}R =0
\ee
where indices $A$ and $B$ go over all of the five coordinates.
One of the homogeneous and isotropic solutions of these vacuum Einstein's equations found in \cite{PonceDeLeon:1988rg} (see also \cite{Overduin:1998pn} for a review) deserves special attention:
\be
ds^2 = dt^2 - e^{2\sqrt{\Lambda/3}\, t} \left(dr^2+r^2d\Omega^2 \right) - d\psi^2
\ee
where $\Lambda = 3/\psi^2$. This metric reduces on $\psi =$ constant hypersurfaces to a $3+1$ dimensional de Sitter metric with $\Lambda =$ constant. An observer located on a $\psi =$ constant slice of the spacetime will measure the effective stress energy tensor
\be
8\pi G T_{\mu\nu} = G_{\mu\nu} \, ,
\ee
where the Greek indices go over $3+1$ dimensional hypersurface only. Note that these equations are already contained in $G_{AB} =0$. Thus, the "matter" described by $T_{\mu\nu}$ is a manifestation of pure geometry in the higher dimensional spacetime and it was called "induced matter" in  \cite{Overduin:1998pn} or "shadow matter" in \cite{Frolov:2003mc,Frolov:2004bq,Stojkovic:2004hp}. The equation of state of matter defined with this stress energy tensor is $p=-\rho$ with $\rho = \Lambda/(8\pi G)$. This solution would not be of much use in theories with compact \cite{add,Starkman:2001xu,Starkman:2000dy} or warped \cite{rs} extra dimensions where the effective size of extra dimensions is small, or brane world models with non-zero tension  where $T_{\mu \nu}$ would get extra contributions. However, in our framework, where the three-dimensional sheet we are located on is embedded into a large structure which is effectively $4+1$ dimensional on cosmological distances, full advantage of this solution can be used. In this framework, we can address the question of the smallness of the observed cosmological constant in a completely different way. We see that observers located at different slices of five-dimensional spacetime infer different values of the effective cosmological constant. The small value of the cosmological constant that we observe can be attributed to the position of our $3+1$ dimensional slice in the full $4+1$ dimensional bulk. Along these lines one might argue that small values of $\Lambda$ (i.e. large values of $\psi$) are much more natural than the opposite. Indeed, the vacuum energy density of $\rho = (10^{-3}$eV)$^4$ corresponds to
the numerical value of $\psi \sim 10^{60} M_{Pl}^{-1}$. This is comparable to the current horizon size, which is in this scenario  comparable to the characteristic distance between $3+1$ dimensional sheets comprising a $4+1$ dimensional structure we live in.

\section{Cosmic microwave background radiation}

Cosmic microwave background radiation in one of the most powerful probes of our universe on all scales. In inflationary scenario, our physical universe is much larger than the current horizon size. Thus, metric perturbations in early universe sample wavelengths exponentially larger than the horizon size in the early universe. In the ordered lattice model (Fig.~\ref{lattice}), this can lead to CMB signatures of geometric properties of the universe larger than the causal horizon at the time of last scattering at a redshift of $z = 1100$\cite{withKinney}. If we have a low-scale inflation which happened after the $2D \rightarrow 3D$ dimensional crossover, then the background for perturbations is effectively three-dimensional, and we can apply standard cosmological perturbation theory. However, the longest wavelength perturbations, {\it i.e.} those which exit the horizon earliest during inflation, will probe length scales of the order of the horizon size in the current universe. Thus, $3D \rightarrow 4D$ crossover might be imprinted into CMB (on angular scales corresponding to the horizon size at the onset of the late-time transition to higher dimension). Taking the angular scale of the current horizon size to be roughly the quadrupole, $\ell_Q = 2$, CMB anomalies resulting from a dimensional transition at redshift $z$ will be present at multipoles larger than a cutoff $\ell_0 \sim \left(1 + z\right)^{3/2} \ell_Q$, which results in the remarkable coincidence that a transition from matter-domination to dark energy-domination at a redshift of $z = 1\ - \ 2$ will correspond to CMB multipoles of $\ell = 5\ - \ 10$, roughly the scales that such anomalies are actually observed in the CMB by WMAP and Planck. Moreover, the basic features of the observed anomalies are consistent with a qualitative expectation from a dimensional transition, i.e. suppressed power due to the presence of standing waves bounded by the intersections of 3D surfaces in the 4D lattice, and mode alignment due to confinement to a lower-dimensional space.

In the opposite limit, CMB might carry infirmation about the shortest scales in our universe. Metric perturbations originate in vacuum fluctuations on scales exponentially smaller than the horizon size during inflation, which in most inflationary models are smaller than the Planck length. This will lead to ``trans-Planckian'' modifications to the primordial power spectrum \cite{Martin:2000xs,Niemeyer:2000eh,Easther:2001fi,Hui:2001ce,Easther:2001fz,Brandenberger:2002hs,Easther:2002xe,Burgess:2002ub,Martin:2003kp,Schalm:2004qk}. The signatures of physics at the quantum gravity scale redshift with the quantum modes during inflation, and are ``frozen'' in the primordial power spectrum at large scales, which might be observable signatures in the CMB. Therefore, signatures of dimensional reduction at very short length scale could be probed by inflationary perturbations, resulting in observable signatures in the CMB. Again assuming that inflation takes place at a low enough scale so that the background is three-dimensional, quantum modes would potentially begin in a vacuum state in a 1D space, transition to 2D, and then to 3D before exiting the horizon. Preliminary studies of this effect have recently been done \cite{Rinaldi:2010yp}.

 Alternatively, inflation itself can take place at a high enough energy scale that the background during inflation is lower dimensional. In this case, the effect on perturbation modes is potentially dramatic, since mode freezing at superhorizon scales would occur in a reduced-dimension background. Finally, it may happen that, due to the peculiarities of lower dimensional physics,  inflation is no longer needed as a solution to the common cosmological problems.

\section{ Running couplings and asymptotic safety}

General Relativity  has had outstanding success in describing all
physical phenomena on large distance (e.g. solar system)
scales. Nonetheless, there is a general consensus that it does not fit
the present paradigm for a fundamental theory of nature, that of a
renormalizable quantum field theory (QFT).  At energy scales much
lower than the Planck mass, $M_{\rm Pl}= G^{-1/2} \simeq 1.22 \times
10^{19}~{\rm GeV}$, GR is described by the Einstein-Hilbert
Lagrangian, $ {\cal L} = - \sqrt{-g} R/16 \pi G$, where $R$ is the
curvature scalar, $g= {\rm det}[g_{\mu \nu}]$, and $g_{\mu \nu}$ is
the metric tensor. The Ricci scalar involves two derivatives acting on
the gravitational field (i.e., the metric $g_{\mu \nu}$).  In an
interaction, each derivative becomes a factor of the momentum transfer
involved, $Q,$ or else of the inverse distance scale probed by the
interaction, $Q \sim 1 / r$.  This implies that $R$ is of order $Q^2$
and therefore the effective strength of the gravitational coupling,
measured by the dimensionless parameter $\sqrt{\tilde G} = Q
\sqrt{G}$, grows without bound if $Q \to\infty$. Of course, $G$, as
any coupling constant in a QFT is subject to renormalization group
(RG) flow. Thus, it is conceivable that Newton's constant has an UV
fixed point (FP), such that if $Q\to\infty$, $G(Q)\sim Q^{-2}$ and
$\tilde G$ will stop growing and tend to a finite limit.

More generally, the effective action of gravitation, derived from
invariance under general coordinate transformations, takes the form
\begin{eqnarray}
S_{\rm eff} & = &  - \int d^4x \, \sqrt{-g} \ \bigg[ g_0(\Lambda)  +
g_1 (\Lambda) R + g_{2} (\Lambda) \ R^2 + g_{3} (\Lambda) \ R_{\mu \nu} R^{\mu \nu} + g_{4} (\Lambda) \ R^3  \nonumber \\
 & &  +  g_{5}(\Lambda)\ R_{\mu \nu} R^{\nu \alpha} R^\mu_\alpha + \dots \bigg] \, ,
\label{effective-action}
\end{eqnarray}
where $R_{\mu \nu}$ is the Ricci tensor, $\Lambda$ is the UV cutoff,
and $g_i(\Lambda)$ are coupling parameters with a cutoff dependence
chosen so that physical quantities are cutoff independent. We have
mentioned that terms of order $R^2,\, R^3,\, \dots$ are negligible at
large enough distances of common experience (e.g., $|g_2|, |g_3| \leq
10^{74}$~\cite{Stelle:1977ry}), but when $\Lambda\to \infty$,
divergences appear in $g_2$ and $g_3$ ($g_4$ and $g_5$) at one (two)
loop(s). We can replace the full set of all renormalized coupling
parameters $g_i(\Lambda)$, with canonical mass dimension $d_i$, by
dimensionless couplings $\tilde g_i(\Lambda) = \Lambda^{-d_i} \, g_i
(\Lambda)$. Because dimensionless, these couplings must satisfy a RG
equation of the form
\begin{equation}
\Lambda \ \frac{d }{d \Lambda}\tilde{g_i} (\Lambda)  = \beta_i \left(\tilde g(\Lambda)\phantom{^i} \!\!\right) .
\end{equation}
The dimensionless couplings can be protected from blowing up if they
are attracted to a finite value. This is known as {\em asymptotic
  safety}~\cite{Weinberg:1}. We say that the theory has a FP if all
the dimensionless couplings tend to finite values $g_i^*$ in the UV
limit. This specific RG behavior guarantees that the theory has a
sensible UV limit, because the FP regime is characterized by the fact
that every dimensionful quantity will scale with $\Lambda$ exactly as
required by its canonical dimension.  For {\em asymptotic safety} to
be possible, it is necessary that $\beta_i(g_i^*) = 0$. It is also
necessary that the physical couplings should be on a trajectory that
is attracted to $g_i^*$. The number of independent coupling parameters
equals the dimensionality of the UV {\em critical surface}, formed by
the locus of points that are attracted towards the FP.  It is
noteworthy that the initial conditions for the flow would be arbitrary
if every trajectory in the space of all couplings has this good
asymptotic behavior. In such a case, all the couplings would have to
be determined by comparison with experimental data and the theory
would be as unpredictive as a nonrenormalizable theory.  Therefore, we
have to require that only a finite number of parameters is left free
by the condition of having a good behavior when $\Lambda \to \infty$,
yielding a finite-dimensional UV critical surface.

Recall that $R$ contains two derivatives of the metric, so in 2D, the
coupling $G$ must be dimensionless and the action
\begin{equation}
S = \int d^2x \, \sqrt{-g} \ \frac{R}{16 \, \pi \, G} = \chi
\end{equation}
is a topological invariant that gives no dynamics to the 2D
metric. The quantity $\chi = 2\, (1-g)$ is known as the Euler
characteristic of the Riemann surface of genus $g$.  In
$(2+\epsilon)$D, $G$ has canonical mass dimension
$-\epsilon$. Defining $\tilde G = G \Lambda^\epsilon$, its beta
function, $\beta_{\tilde G} = \epsilon \tilde G - {\cal B} \tilde
G^2,$ has a FP at $\tilde G = \epsilon/{\cal B}$, where ${\cal B} =
\frac{38}{3}$~\cite{Weinberg:2}. Therefore, for sufficiently small
$\epsilon$, there is a $(2 + \epsilon)$D {\em asymptotically safe}
theory of pure gravity, with a one-dimensional critical surface. By
adding matter fields with a minimal coupling to gravity one obtains
${\cal B} = \frac{38}{3} + 4N_v - \frac{1}{3} N_f - \frac{2}{3} N_s,$
where $N_s$, $N_v$, and $N_f$ are the number of scalar, vectors, and
Majorana fermion fields, respectively~\cite{Gastmans:1977ad}.  Hence,
{\em asymptotic safety} would still be preserved provided that there
are enough gauge fields to balance any scalar or fermion fields and
that the couplings of the matter fields with themselves do not raise
problems.

Dimensional continuation from $(2+\epsilon)$ to 4D is driven by the
truncated exact RG equation~\cite{Reuter:1996cp}. Here, the effective
average action, $\Gamma[g,\Lambda]$, is written as a sum of a finite
number of terms -- like those shown explicitly in
Eq.~(\ref{effective-action}) -- and ignoring the fact that the {\em
  beta functional} inevitably does not vanish for the couplings of
other terms in the functional $\Gamma[g,\Lambda]$, which in a given
truncation are assumed to vanish.  Arguably, it seems encouraging that
by considering a Lagrangian containing all terms of the form $R^0, R^1
\dots R^6$, for which the space of coupling constant is of dimension
7, one obtains a 3-dimensional UV critical surface (i.e., the number
of independent couplings is equal to
3)~\cite{Codello:2006in}. Surprisingly, the RG flow predicts an
effective dimensionality of spacetime which is scale dependent: it
equals 4 at macroscopic distances, but gets dynamically reduced at
short distances and spacetime becomes a 2D
fractal~\cite{Lauscher:2001ya}.

Another seemingly different, but perhaps closely related subject is
the Regge regime of QCD~\cite{Cheng:1987ga}. It has indeed been
suspected for some time that there exists an intimate relationship
between QCD at high energies and two dimensional field
theory~\cite{Lipatov:1988ce}.  Efforts have been made to obtain an
appropriate scheme to study QCD in the Regge regime of large energies
$\sqrt{s} \to \infty$ and fixed momentum transfers $Q(Q^2 = -t)$ ($|Q|
\sim 1~{\rm GeV}$, i.e., $|Q| \gg \Lambda_{\rm QCD} \sim 100~{\rm
  MeV}$). It was recognized~\cite{Nachtmann:1991ua} that in this
regime neither QCD perturbation theory applies --since $|t|$ is too
small,-- nor can the usual lattice gauge theory approach give
numerical answers directly -- since $s$ is too large. Formally, QCD
scattering in the near-forward limit can be understood as the mixing
of a ``short-distance'' phenomenon in the longitudinal coordinates and
a ``long-distance'' phenomenon in the transverse coordinates. In
particular within the framework of perturbation theory, systematic
procedures have developed for extracting the large $s$, fixed $t$,
behavior of each amplitude and for summing these contributions using
the leading logarithmic or eikonal approximation scheme. It is a
striking fact that by treating the longitudinal and transverse degrees
of freedom separately, the contribution at each order takes the form
of two dimensional amplitudes~\cite{Lipatov:1988ce,Verlinde:1993te}.

\section{Causal dynamical triangulation}

Another concrete realization of the dimensional reduction at high energies is known as
the causal dynamical triangulation~\cite{Ambjorn:1998xu}.
This alternative approach to quantum gravity is based on the path
integral formulation of Euclidean gravity~\cite{Hawking:1978jz}. In
the regularization scheme of dynamical triangulations, the functional
integral over Euclidean metrics,
\begin{equation}
\int {\cal D} [g] \, e^{-S_{\rm E}[g]} \, ,
\end{equation}
is discrete ($S_{\rm E}[g]$ is the Euclidean Einstein-Hilbert
action). The sum is taken over all possible manifold-gluings of a set
of equilateral simplicial building blocks~\cite{David:1984tx}. (Gluing
together triangles at their edges leads to a 2D surface and gluing
four-simpleces along their faces -- which are actually
three-dimensional tetrahedra -- can produced a 4D manifold.)

However, such a non-perturbative superposition of 4D universes is inherently
unstable: quantum fluctuations of curvature in short scales, which
characterize the different superposed universes contributing to the
average, do not cancel one another out to produce a smooth, classical
universe on large scales. The problem in assembling individual
universes is apparently in the roots of Euclidean quantum gravity,
which does not build in a notion of causality. To enforce an arrow of
time in the ``gluing rules'' one needs to use causal dynamical
triangulations. The causality conditions imposed in the Lorentzian
dynamically triangulated gravitational path integral act as an
effective regulator on the geometry, still allowing for large
curvature fluctuations, but suppressing changes in the spatial
topology~\cite{Ambjorn:2000dv}. The superposition of all possible
Lorentzian spacetime shapes also yields a dynamical fractal structure,
with a short-distance spectral dimension that increases smoothly from
$D_s = 1.8 \pm 0.25$ to an asymptotic value $D_s = 4.02 \pm
0.1$~\cite{Ambjorn:2005db}. It is therefore not unlikely that the
mechanism of a dynamical dimensional reduction from 4 to 2 dimensions
in asymptotically safe gravity is the same phenomenon as the
dimensional reduction observed in the Monte Carlo studies of causal
dynamical triangulations~\cite{Lauscher:2005qz}.

\section{Other models of dimensional reduction}

The results from asymptotic safety and causal dynamical
triangulations seem to point to the correct direction of a spacetime
with evolving dimensionality. Additional theoretical evidence for dimensional reduction at high energies comes
from the strong-coupling limit of the Wheeler-DeWitt equation \cite{Carlip:2009km} and phenomenon of ``asymptotic silence" \cite{Carlip:2011tt}.
It has also been shown that a non-commutative quantum spacetime with minimal length scale will exhibit the properties of a two-dimensional manifold  \cite{lmpn1}. Reducing the number of dimensions in the far UV limit offers a completely new approach to gauge couplings unification \cite{Shirkov:2010sh}.
Applications of models with dimensional reduction to some fundamental problems can be further found in
\cite{Calcagni:2009kc,Calcagni:2010bj,Nicolini:2011nz,Calcagni:2011sz,Modesto:2011kw,Obukhov:2011ks,Mann:2011rh,Nieves:2011fy,Mureika:2011py,Landsberg:2010zz,Stoica:2013wx,GonzalezMestres:2010pi,Calmet:2010vp,Caravelli:2010be}.

\section{Experimental evidence}

 On the experimental front, it is very intriguing that some evidence for the lower dimensional structure of our spacetime on a TeV scale might already exist. Namely,
alignment of the main energy fluxes in a target (transverse) plane has been observed in families of cosmic ray particles in high altitude cosmic ray experiments \cite{et1986,Mukhamedshin:2005nr,Antoni:2005ce} (high altitude is crucial in order to catch the very beginning of the shower before the energies significantly degrade).
First, an intriguing alignment of gamma-hadron families (i.e., the outgoing
high energy secondary particles from a single collision in the
atmosphere) along a straight line in a transverse plane has
been observed with (lead and carbon) $X$-ray emulsion chambers
($X$REC's) in the Pamir mountains~\cite{et1986}.

[The
  Pb-chambers are assembled of many sheets of lead (1~cm thick)
  interlaid with $X$-ray films. This provides a few interaction
  lengths for hadrons and a quasicalorimeter determination of the
  particle's energy. The C-chambers contain a 60~cm carbon layer
  covered on both sides by lead plates sandwiched with $X$-ray
  films. The carbon block provides a large cross section for hadron
  interaction, while the lead blocks are of minimal thickness allowing
  determination of particle energies. The total area of the chambers
  is few tens of square meters. Electron-photon cascades initiated by
  high energy hadrons and gamma-rays inside the $X$REC's produced dark
  spots whose sizes are proportional to the cascade energy deposited
  on the $X$-ray film.]

  These families can be reconstructed by
measuring the coordinates and the incident direction of each particle
in the film emulsion. This allows determination of the total energy in
gamma-rays and the total energy of hadrons release to
gamma-rays. Recall that most of the hadrons in the family are pions
and the average fraction of energy transferred by pions to the
electromagnetic component is $\simeq 1/3$.  All families in the
experiment are classified by the value of the total energy observed in
gamma-rays, $\sum E_\gamma$. The centers of the main energy fluxes
deposited on the $X$-ray film (a.k.a. ``subcores'') include halos of
electromagnetic origin, gamma-ray clusters, single gamma-rays of high
energy, and high energy hadrons. The criterion for alignment is given
by the asymmetry parameter
\begin{equation}
\lambda_N = \frac{1}{N (N-1) (N-2)}\sum_{i\neq j \neq k} \cos 2 \varphi_{ij}^k \,,
\end{equation}
where $N$ is the number of subcores and $\varphi_{ij}^k$ is the angle
between vectors issuing from the $k$-th subcore to the $i$-th and
$j$-th subcores~\cite{Ivanenko:1992qw}. The parameter $\lambda_N$
decreases from $1$ (corresponding to $N$ subcores disposed along a
straight line) to $-1/(N-1)$ (corresponding to the isotropic case).
Events are referred to as aligned if the $N$ most energetic subcores
satisfy $\lambda_N \geq \lambda_N^{\rm cut}$. A common choice is $N=4$
and $\lambda_N^{\rm cut} = 0.8$.

The data have been collected at an altitude of 4400~m a.s.l., {\it
  i.e.}, at a depth of 594~g/cm$^2$ in the atmosphere. For low energy
showers, $30~{\rm TeV} < \sum E_\gamma < 200~{\rm TeV}$, the
fraction of aligned events coincides with background expectation from
fluctuations in cosmic ray cascade developments. However, for $\sum
E_\gamma > 700~{\rm TeV}$, the alignment phenomenon appears to be
statistically significant~\cite{Mukhamedshin:2005nr}.  Namely, the
fraction ($f$) of aligned events is $f(\lambda_4 \geq 0.8) = 0.43 \pm
0.17$ (6 out of 14) in the Pb-$X$REC catalogue, and $f(\lambda_4 \geq
0.8) = 0.22 \pm 0.05$ (13 out of 59) in the C-$X$REC catalogue. The
predominant part of the gamma-hadron families is produced by hadrons
with energy $E_0 > 10 \sum E_\gamma$, corresponding to interactions
with a center-of-mass energy $\sqrt{s} > 4$~TeV.  Data analyses
suggest that the production of most aligned groups occurs low above
the chamber~\cite{Ivanenko:1992qw}. Thus, it is not completely
surprising that the KASCADE Collaboration has found no evidence of
this intricate phenomenon at sea level ($\sim
1000$~g/cm$^2$)~\cite{Antoni:2005ce}.

Interestingly, the fraction of events with alignment registered in
Fe-$X$REC's at Mt. Kanbala (in China) is also unexpectedly
large~\cite{Xue:1999bb}. For gamma-hadron families with energy $\sum
E_\gamma \geq 500$~TeV the fraction of aligned events is $f(\lambda_3
\geq 0.8) = 0.5 \pm 0.3$ (3 out of 6). In addition, two events with
$\sum E_\gamma \geq 1000$~TeV have been observed in stratospheric
experiments~\cite{Capdevielle:1988pe}. Both events are highly aligned:
{\it (i)} the so-called STRANA superfamily, detected by an emulsion
chamber on board a Russian stratospheric balloon, has $\lambda_4 =
0.99$; {\it (ii)} the JF2af2 superfamily, detected by an emulsion
chamber during a high-altitude flight of the supersonic aircraft
Concord, has $\lambda_4 = 0.998$. It is worth noting that
stratospheric experiments record the alignment of particles, whereas
mountain-based facilities register the alignment of the main fluxes
of energy originated by these particles on a target plane.

The strong collinearity of shower cores has been interpreted as a
tendency for coplanar scattering and quasiscaling spectrum of
secondary particles in the fragmentation
region~\cite{Deile:2010mv}. If the aligned phenomenon observed in
cosmic ray showers is not a statistical fluctuation, then events with
unusual topology may be produced at the Large Hadron Collider (LHC).
Lower dimensional scattering has very important predictions for the Large Hadron Collider physics.  There are three consequences which should be observable if the physics becomes planar at the TeV scale: {\it (i)\/} cross-section of hard scattering processes changes compared to that in the SM as the momentum transfer becomes comparable with the crossover scale; {\it (ii)\/} $2 \to 4$ and higher order scattering processes at high energies become planar, resulting, {\it e.g.\/}, in four-jet events, where all jets are produced in one plane in their center-of-mass frame, thus strikingly different from standard QCD multijet events; {\it (iii)\/} under certain conditions, jets of sufficiently high energy may become elliptic in shape (for details see \cite{Anchordoqui:2010er,Anchordoqui:2010hi,Stojkovic:2013lga}). It is also important to note that (in the ordered lattice model) no new fundamental particles are expected to exist in order to solve the hierarchy problem.

\section{Gravity waves}

A distinct prediction of a dimensional reduction scheme comes from the nature of gravity in lower dimensions.  It is well-known that, in a $(2+1)$-dimensional general relativity, there are no local gravitational degrees of freedom, and hence there are no gravitational waves (or gravitons).  If the universe was indeed $(2+1)$-dimensional at some earlier epoch, it is reasonable to deduce that no primordial gravitational waves of this era exist today.  There is thus a maximum frequency of primordial gravitational waves, implicitly related to the dimensional transition scale, beyond which no waves can exist.  This indicates that gravitational wave astronomy can be used as a tool for probing this scale \cite{Mureika:2011bv}. If $2D \rightarrow 3D$ dimensional crossover happened when the temperature in the universe was around $1$TeV, then the Laser Interferometer Space Antenna (LISA) should be able to register the cut-off frequency beyond which there are no gravity waves. However, due to NASA budgetary constraints, the original mission was canceled,
and European-led ESA is considering a scaled-down variation on the original LISA mission temporarily
named the New Gravitational-Wave Observatory (NGO).
The originally proposed LISA mission planned to deploy three spacecraft in a triangular constellation with $5 \times 10^9~$m side lengths orbiting
the Sun, $20^\circ  $ behind the Earth. The baseline NGO configuration is a triangular constellation
with $1 \times 10^9~$m arms located $9^\circ $ behind the Earth in orbital phase. Thus, the NGO gravitational-wave detector will
have shorter arms, and as a result its sensitivity will be peaked at higher frequencies.
Based on the previous estimates, LISA sensitivity was sufficient to successfully probe our universe at temperatures of TeV and higher. Since the sensitivity of the detector scales down with the square root of the length of the of the interferometer arm, if the length is reduced by the factor of five, sensitivity would would go down only by a factor of two. This leaves a lot of room for an optimism that an eventual dimensional cross-over could be detected even with significantly scaled-down missions.

While GR does not admit gravitational waves in spacetimes lower than four dimensions, many known exotic or alternative theories of gravity do.  These include Horava-Lifshitz gravity \cite{horava}, massive graviton theories \cite{massive}, and others. In possible extensions and variations of the evolving dimensions scenario where the lower dimensional gravity significantly differs from GR, the dimensional cross-over will also be clearly marked by the change in nature of gravity waves. It will be interesting to calculate the details of the gravity wave signature of the dimensional transition in these models.

\section{Concrete model of evolving dimensions: Stringy Model}

From the model building point of view,  a framework of ``evolving" or ``vanishing" dimensions was proposed in \cite{Anchordoqui:2010er} in which the space at scales shorter than $10^{-17}$ cm is lower dimensional, while at scales larger than a Gpc it is higher dimensional. In this setup, the number of dimensions increases with the length scale.  On the shortest distances at which our space appears as continuum, the space is one-dimensional. At a certain critical length scale, the space becomes effectively two-dimensional. At the scale of about $10^{-17}$ cm, the space becomes effectively three-dimensional. Finally, at the scales of about a Gpc, the space becomes effectively four-dimensional. In a dynamical picture where the universe starts from zero size and then grows, the dimensions open up as the universe expands and temperature drops. The {\it ad hoc} model that was used in this proposal was an ordered lattice, which captures all the basic features of the proposal and allows one to make generic model independent predictions. However, so far no explicit model in terms of fundamental Lagrangians was constructed.

In this section, we outline a string theory inspired explicit model of ``evolving dimensions" \cite{withNiayesh}. To do so, we will use the existing apparatus of the string theory, where dimensions are viewed as fields, and modify it to achieve the change of dimensionality of our space with the energy scale.

We can start from the standard Nambu-Goto action:
\be \label{nga}
S_{\rm free} = -\frac{1}{2 \pi \alpha '} \int d^2 \xi \sqrt{-\gamma}
\ee
where $\alpha'^{-1}$ is the string tension, and $\gamma$ is the determinant of the metric on the string world sheet  $\gamma_{ab}$
\be \label{gamma}
\gamma_{ab} = g_{\mu \nu} \partial_a X^\mu \partial_b X^\nu .
\ee
The metric in the target space $g_{\mu \nu}$ is usually considered to be fundamental and  $\gamma_{ab}$ induced. However, we will adopt the opposite view here. The lower dimensional metric $\gamma_{ab}$ will be considered fundamental, and higher dimensional manifold $g_{\mu \nu}$ induced since it is woven by an evolving lower dimensional submanifold (as in Fig.~\ref{1d2d}).
Coordinates on the string world-sheet are $\xi^a = (\tau, \sigma)$. The coordinates in the target space are $X^{\mu}(\tau, \sigma)$. The index $\mu = (0,1,2,3, ... , n)$ where $n$ is the number of dimensions in the target space. $X^{\mu}$ represent the coordinates in the space that we live in (we do not fix the dimensionality of that space {\it a priori}).
We can also understand $X^{\mu}$ as fields that live on the string world-sheet, so in principle, we can add mass terms for them. This would break the conformal symmetry, which would require more careful standard string theory interpretation. However, for the purpose of a phenomenological theory with desired properties, conformal symmetry is not crucial.

We will now illustrate the basic idea by adding temperature dependent masses for the fields $X^{\mu}$
\bea \label{lmass}
&& L_{\rm mass} = \sum_{i=1}^{n} M(m_i,T)^2 X^i X_i = \\ && m_0^2 e^{-m_1/T} X^1 X_1 +
 m_0^2e^{-m_2/T} X^2 X_2 + \nonumber \\ && m_0^2e^{-m_3/T} X^3 X_3 + \ldots + m_0^2e^{-m_n/T} X^n X_n  \nonumber
\eea
where $m_1 \gg  m_2 \gg m_3 \gg \ldots \gg m_n$. Note that the time-like coordinate $X^0$ is massless, so it is always excited (this assumption can easily be changed). Parameter $m_0$ has units of mass, and is of the order of the fundamental energy scale (perhaps $M_{Pl}$). We start from a hot Big Bang when $T \gg m_1, m_2, m_3, \ldots , m_n$, so all the fields are massive and are not excited. When the temperature drops to $T \ll m_1$, only the first field $X^1$ is practically massless and gets excited; when the temperature drops to $T \ll m_2$, only  $X^1$ and $X^2$ are excited, and so on. Today, at $T \sim 10^{-3}eV \ll m_3$, the first three fields are excited. In principle, the process does not have to stop at any finite number, so decrease in energy would be opening more and more dimensions. Note that we can always use the gauge freedom to identify the time coordinate on the worldsheet with the time coordinate in the target space, which will result in identifying the temperature on the worldsheet with that in the target space. Thus the fundamental field(s) $\phi(\xi^a)$ which live on the worldsheet actually produce effective temperature in the target space.

The temperature dependent mass terms, similar to those in Eq.~(\ref{lmass}) can generically be introduced via non-perturbative IR effects. For example, photons in a thermal electron-positron plasma at $T \lesssim m_e$ have a mass given by the plasma frequency:
\be
m_{\gamma}^2 = \frac{8\pi n_e \alpha_e}{m_e} = \left( 128 T^3 m_e\over \pi \right)^{1/2} \alpha_e e^{-m_e/T},
\ee
where $\alpha_e$ is the fine structure constant and $m_e$ is the electron mass. In this case, the collective interaction of photons with the plasma induces an effective mass, which becomes negligible at low temperatures, as the density of thermally produced electron-positron plasma vanishes exponentially at $T \ll m_e$.

The action corresponding to the mass Lagrangian (\ref{lmass}) is
\be \label{smass}
S_{\rm mass} = \frac{1}{2 \pi \alpha '} \int d^2 \xi \sqrt{-\gamma} L_{\rm mass}
\ee
Note that fields $X^\mu$ have the physical interpretation as dimensions and have units of length, which makes the action (\ref{smass}) dimensionless.

The equations of motion that govern the excitations of the fields $X^\mu$ are
\bea  \label{eom1}
&& \left( \Box +   m_0^2 e^{-m_1/T} \right) X^1  = 0 \\
&& \left( \Box +   m_0^2 e^{-m_2/T} \right) X^2 = 0 \nonumber  \\
&& \ldots  \nonumber
\eea
where $\Box =\frac{1}{\sqrt{-\gamma}} \partial_a \left(\sqrt{-\gamma} \gamma^{ab} \partial_b   \right)$.

Since all the components of the field multiplet $X^\mu$ have different masses, the general Lorentz (and diffeomorphism) invariance in the target space is broken. However, at any given temperature the Lorentz invariance is restored in some subset of the original $n$ dimensions, e.g. today at $T\sim 10^{-3}$eV the Lorentz invariance is effectively restored in the first three dimensions since all the first three fields are practically massless. This also implies that the usual string theory conformal invariance is restored in the subset of dimensions which are practically massless. This is important if one would like to preserve the straightforward string theory interpretation (which must include massless gravitons and gauge fields in its spectrum).

The concrete action $S=S_{\rm free} + S_{\rm mass}$ allows us to address explicitly the problems in cosmology, gravity and high energy physics. For any concrete problem, a particular metric  $g_{\mu \nu}$ should be inserted in the action (\ref{nga}). For example, for black hole physics in the evolving background, $g_{\mu \nu}$ would be the Schwarzschild metric;  for cosmological problems in early and late universe, $g_{\mu \nu}$ would be the FRW or de-Sitter metric.

Depending on the relevant scales in this model, excitations of the fields $X^\mu$ that we call dimensions may behave as a rigid frustrated string network in one limit, or densely fluctuating space-filling structure in the opposite limit (or something in between).
For a finite tension string, the classical non-fluctuating string network as in Fig.~\ref{1d2d}, has a finite dimensional crossover scale. If quantum fluctuations are giving only small and negligible corrections to this picture, we practically have the ordered lattice model of vanishing dimensions introduced in \cite{Anchordoqui:2010er}. In such a model, the lattice is rigid and classical, and propagation of particles along the lattice links/blocks depends on the particle energy, i.e. short wavelength particles see $1D$ string, while large wavelength particles see a $2D$ surface.
However, quantum fluctuations could be significant for a small tension string. Fluctuations smear the string and can even make it space-filling, in which case a $1D$ string would be densely covering a $2D$ space. Since the mass terms in Eq.~(\ref{lmass}) depend on temperature, fluctuations would depend on the temperature of the environment (apart from the string tension). Therefore, the number of dimensions that a particle can see would strongly depend on the temperature of the environment, not energy of the particle.

These two extreme limits (classical rigid network and space-filling string) in fact provide quite different phenomenology. The ordered lattice model implies that the number of dimensions that an individual particle sees changes with energy (wavelength) of that particle, what could be in principle tested in high energy particle collisions (like planar events, elliptic jets etc). In contrast, Eq.~(\ref{lmass}) implies that the number of dimensions changes with temperature, which requires finite energy within some finite volume. Thus, in the limit of the space-filling string, in order to de-excite dimensions one has to raise energy in some finite region of space. This can be achieved in heavy ion collisions where multiple particles collide, but not in two-particle collisions. An obvious problem is then that one has to raise the energy by four orders of magnitude in order to raise the temperature by one order of magnitude (since $E/V \sim T^4$), which would make the $3D \rightarrow 2D$ dimensional cross-over invisible at the LHC even if the crossover temperature is as low as $1$ TeV, which has deep implications for the new physics at the LHC.  A clear prediction of this limit of temperature dependent number of dimensions is that the LHC will be practically blind to new physics even when working with its full power. Only slight deviations from the standard $(3+1)$-dimensional physics might be expected. A possible theoretical drawback of this extreme scenario is that the hierarchy problem might not be solved by temperature dependent physics, since the corrections to the Higgs mass will still be quadratically divergent in vacuum. In contrast, the classical rigid network (or ordered lattice) limit would clearly solve the hierarchy problem since high energy particles in vacuum propagate in lower dimensional spacetime.

Of course, a generic situation is somewhere between these two extreme limits of energy vs. temperature dependent new physics. A smooth transition between these two limits may be observed in events with high multiplicity and also density (number of particles per unit volume).
The best place to look for experimental evidence are the cosmic rays experiments. Cosmic rays collide particles in our atmosphere with center of mass energies of $100$TeV and more, which is high above the LHC energies. What is even more important, cosmic rays often produce very high multiplicity events with hundreds of particles in the single collision. Though it is very difficult to determine whether full thermal equilibrium had been established during the interaction, this regime is much closer to the high temperature environment than the events at the LHC. This might be the main reason why the earlier high altitude cosmic rays experiments observed planar propagation of secondary showers \cite{et1986,Mukhamedshin:2005nr,Antoni:2005ce}. It was noticed that only super-families with very high number of particles have planar alignment. The problem there was that very few super-families were observed, so it is still not clear if this effect was a statistical fluke or not. Current cosmic rays experiments are not performed at high altitude, so it seems very unlikely to replicate the results since energy of the shower degrades very quickly if one is not able to catch the very beginning of the shower at a high altitude.
The only exception here might be neutrinos.  Neutrinos interact weakly and unlike protons and photons can penetrate the whole depth of the atmosphere and interact for the first time in the detector so that the beginning of the shower can be caught.  Indeed, IceCube recently detected two PeV neutrino events which light up the whole detector by producing hundreds of particles \cite{Aartsen:2013bka}. Unfortunately, these events had the center of mass energy of only $1.4$ TeV, while the observed threshold for the planar events in earlier experiments was $4$TeV.  It will be very important to collect events that originate in the detector and have above the threshold energy, and check the topology of the produced showers. If earlier observed alignment is also observed by IceCube, this might strongly support the model we discussed here.

A different concrete realization of ``vanishing" or ``evolving dimensions" in the context of cascading gravity can be found in \cite{Hao:2014tsa}.

\section{Potential problem: Lorentz invariance violations}

\begin{figure}[htb]
\vspace*{-0.1in}
\center{\scalebox{0.65}{\includegraphics{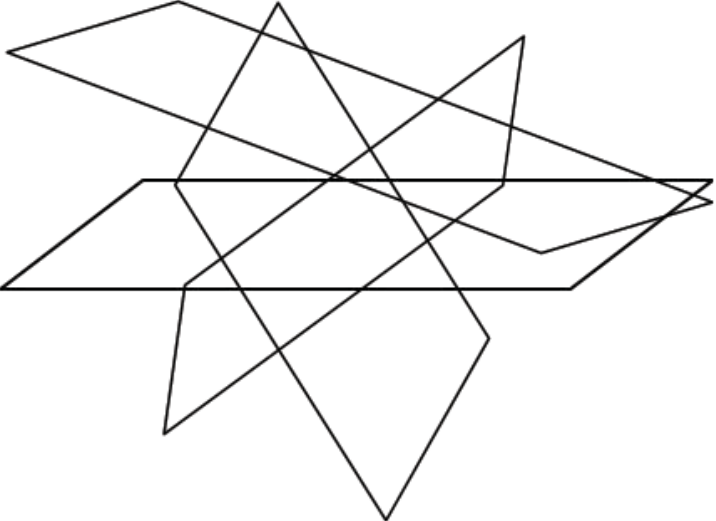}}}
\vspace*{-0.1in}
\caption{Random orientation of lower-dimensional planes may avoid systematic violation of Lorentz invariance.}
\label{random}
\end{figure}

One of the potential problems with this model of emerging dimensions might be possible Lorentz invariance violations in the light of strong Fermi constraints.   High energy photons propagating from a distant part of the universe toward us may be affected by discrete nature of spacetime which in turn could modify the dispersion relation. Concretely, in the discrete ordered lattice limit, photons with the wavelength larger than the dimensional cross-over length scale would propagate in $3+1$-dimensional spacetime, while those with the much shorter wavelength would see $(2+1)$-dimensional spacetime. This may potentially lead to modified dispersion relation, or a time arrival delay when two photons (one above and one below the cross-over scale) are compared. One of the ways to evade strong Lorentz invariance violations is to have a random lattice (as in Fig.~\ref{random}), where Lorentz invariance violation would be stochastic and would average to zero, thus avoiding systematic violation of the dispersion relations. We also note that the two photons used by Fermi to put the constraints were both below TeV energies \cite{Vasileiou:2010nx}. They observed one $3$GeV and one $31$GeV photon coming to the detector with the time delay of about 1 second. However, if the crossover scale is set to a TeV by the solution to the hierarchy problem, none of these two photons actually probes the lower dimensional regime.
Obviously, in the opposite limit of the space-filling densely fluctuating string, individual high energy quanta propagating toward us from the other end of the universe do not propagate in the high temperature regime. For them to see a lower dimensional spacetime, they would have to propagate through hot plasma with temperature higher than the cross-over scale. Thus, they always see $3+1$-dimensional spacetime and Fermi constraints do not affect them at all.

On the other end, if we lower the temperature in some region of space below the current cosmological temperature of $10^{-3}$eV, we could be exciting new dimensions. Then, in order to avoid eventual constraints from cool condensed matter systems in the lab environment (where we can cool particles almost down to absolute zero), one would need to add the coupling of $X^\mu$ to the vacuum energy density (either fundamental or effective). In that case reducing the kinetic energy of the particles in the cool system would not eliminate the vacuum energy density within some volume. The, the next dimensional cross-over ($3+1$ to $4+1$) would happen if and when the total energy density drops below some critical value lower than   $(10^{-3} {\rm eV})^4$.

\section{Planck scale may not be $10^{19}$GeV}

The scale at which quantum effects in gravity become important is known as the Planck scale. In 3D, there is a unique expression with dimensions of mass (energy) that involves all the known fundamental constants
\be
M^{3D}_{\rm Pl} = \sqrt{\frac{\hbar c}{G_{3D}}}
\ee
If we substitute the measured vales for $c$, $\hbar$ and $G_{3D}$, we get the numerical value of $M^{3D}_{\rm Pl} = 10^{19}$GeV. However, this is inherently a three-dimensional value. In the context of large extra dimensions, we learned that  $M_{\rm Pl}$ could be as low as $1$TeV if gravity can propagate in more than three spatial dimensions. In the context of evolving dimensions, the fundamental scale of gravity is much more peculiar. In two dimensions, there is no quantity with dimensions of mass that involves all  the constants $c$, $\hbar$ and $G$. The only quantity that we can construct in $2D$ is
\be
M^{2D}_{\rm Pl} = \frac{c^2}{G_{2D}}
\ee
The absence of $\hbar$ is notable. This may be perhaps connected with the fact that $2D$ general relativity has no local propagating degrees (i.e. no gravitons in quantum case).
Since $G_{2D}$, unlike $G_{3D}$, is a number of unknown magnitude, the exact numerical value of $M^{2D}_{\rm Pl}$ is also unknown. In principle, the value of $M^{2D}_{\rm Pl}$ could be much greater than $10^{19}$GeV.

If we formally extend our discussion to one spatial dimension, we will find that there is no quantity with dimensions of mass that involves any combination of $c$, $\hbar$ and $G$ (or at least some of them). This is perhaps connected with the fact that $1D$ general relativity is not a dynamical theory.

If we accept the assumption that a fundamental high energy theory is lower dimensional, and our $3D$ theories are just low energy approximations, then it appears that general relativity is also an emergent theory that emerges for the first time in $2D$, and becomes a fully fledged propagating theory only in $3D$. In that case $M^{3D}_{\rm Pl} = 10^{19}$GeV plays no fundamental role.  The ultimate theory of space-time has its own fundamental scale, which for all we know could be much lower or much greater than $10^{19}$GeV.
This would also imply that it does not make much sense to quantize gravity as we know it, especially not in $3D$.

\section{Conclusions}

We presented a brief review of models that have one thing in common - reduced dimensionality at high energies. The standard lore in high energy physics so far was to introduce new structures and degrees of freedom in order to solve the long standing problems. However, it appears that going in the opposite direction is more promising. In lower dimensions, many problems that plague our standard theories simply do not exist. Moreover, some experimental evidence for the dimensional reduction may already exist, and future experiments and observations may give us a more definite picture.

{\it Acknowledgement:}
This work was  partially supported by the US National Science Foundation, under Grant No. PHY-1066278 and PHY-1417317.
The author also acknowledges hospitality and support from the Perimeter Institute for Theoretical Physics.


\begin{thebibliography}{99}




\bibitem{Carlip:1995zj}
  S.~Carlip,
  J.\ Korean Phys.\ Soc.\  {\bf 28}, S447 (1995)
  [arXiv:gr-qc/9503024].



\bibitem{Klosch:1997md}
  T.~Klosch and T.~Strobl,
  Class.\ Quant.\ Grav.\  {\bf 14}, 1689 (1997)
  [arXiv:hep-th/9607226].

\bibitem{LouisMartinez:1993cc}
  D.~Louis-Martinez and G.~Kunstatter,
  Phys.\ Rev.\  D {\bf 49}, 5227 (1994).


\bibitem{Grumiller:2006ja}
  D.~Grumiller and R.~Meyer,
  Class.\ Quant.\ Grav.\  {\bf 23}, 6435 (2006)
  [arXiv:hep-th/0607030].

\bibitem{Zaslavskii:2003eu}
  O.~B.~Zaslavskii,
  Class.\ Quant.\ Grav.\  {\bf 20}, 2963 (2003)
  [arXiv:hep-th/0305199].


\bibitem{Giddings:1992ae}
  S.~B.~Giddings and A.~Strominger,
  Phys.\ Rev.\  D {\bf 47}, 2454 (1993)
  [arXiv:hep-th/9207034].




\bibitem{Callan:1992rs}
  C.~G.~.~Callan, S.~B.~Giddings, J.~A.~Harvey and A.~Strominger,
  Phys.\ Rev.\  D {\bf 45}, 1005 (1992)
  [arXiv:hep-th/9111056].


\bibitem{Bogojevic:1998ma}
  A.~Bogojevic and D.~Stojkovic,
  Phys.\ Rev.\  D {\bf 61} (2000) 084011
  [arXiv:gr-qc/9804070].


\bibitem{cdgp} G.R. Dvali, G. Gabadadze, and M. Porrati, ``4D Gravity on a Brane in 5D Minkowski Space ,'' Phys. Lett. {\bf B} 485, 208 (2000); C. de Rham, G.R. Dvali, S. Hofmann, J. Khoury, O. Pujolas, M. Redi, and A.J. Tolley, ``Cascading DGP ,'' Phys. Rev. Lett. {\bf 100}, 251603 (2008);
  R.~Gregory, V.~A.~Rubakov and S.~M.~Sibiryakov,
  Phys.\ Rev.\ Lett.\  {\bf 84}, 5928 (2000)
  [hep-th/0002072].







\bibitem{Anchordoqui:2010er}
  L.~Anchordoqui, D.~C.~Dai, M.~Fairbairn, G.~Landsberg and D.~Stojkovic,
  Mod.\ Phys.\ Lett.\ A {\bf 27}, 1250021 (2012)
  [arXiv:1003.5914 [hep-ph]].





\bibitem{Anchordoqui:2010hi}
  L.~A.~Anchordoqui, D.~C.~Dai, H.~Goldberg, G.~Landsberg, G.~Shaughnessy, D.~Stojkovic and T.~J.~Weiler,
  Phys.\ Rev.\ D {\bf 83}, 114046 (2011)
  [arXiv:1012.1870 [hep-ph]].

\bibitem{Stojkovic:2013lga}
  D.~Stojkovic,
  arXiv:1304.6444 [hep-th].


\bibitem{Mureika:2011bv}
  J.~R.~Mureika and D.~Stojkovic,
  Phys.\ Rev.\ Lett.\  {\bf 106}, 101101 (2011)
  [arXiv:1102.3434 [gr-qc]];
  Phys.\ Rev.\ Lett.\  {\bf 107}, 169002 (2011)
  [arXiv:1109.3506 [gr-qc]].



\bibitem{Li:1994et}
  M.~Li and C.~I.~Tan,
  Phys.\ Rev.\  D {\bf 50}, 1140 (1994)
  [arXiv:hep-th/9401134].

\bibitem{Engelhardt:1995qm}
  M.~Engelhardt and B.~Schreiber,
  Z.\ Phys.\  A {\bf 351}, 71 (1995).


\bibitem{et1986}
   L. T. Baradzei {\it et al.} [Pamir Collaboration],
  Bull.\ Russ.\ Acad.\ Sci.\ Phys.\  {\bf 50N11}, 46 (1986)
  [Izv.\ Ross.\ Akad.\ Nauk Ser.\ Fiz.\  {\bf 50}, 2125 (1986)];
  MGU-89-67-144 (1989);
  Bull.\ Russ.\ Acad.\ Sci.\ Phys.\  {\bf 55N4}, 24 (1991)
  [Izv.\ Ross.\ Akad.\ Nauk Ser.\ Fiz.\  {\bf 55}, 650 (1991)];
  Bull.\ Russ.\ Acad.\ Sci.\ Phys.\  {\bf 57}, 612 (1993)
  [Izv.\ Ross.\ Akad.\ Nauk Ser.\ Fiz.\  {\bf 57N4}, 40 (1993)].




\bibitem{Mukhamedshin:2005nr} R.~A.~Mukhamedshin,  JHEP {\bf 0505}, 049 (2005).

\bibitem{Antoni:2005ce}  T.~Antoni {\it et al.}  [KASCADE Collaboration],  Phys.\ Rev.\  D {\bf 71}, 072002 (2005)  [arXiv:hep-ph/0503218].




\bibitem{Aartsen:2013bka}
  M.~G.~Aartsen {\it et al.}  [IceCube Collaboration],
  arXiv:1304.5356 [astro-ph.HE].

\bibitem{Vasileiou:2010nx}
  V.~Vasileiou [Fermi LAT and GBM Collaborations],
  arXiv:1002.0349 [astro-ph.HE].




\bibitem{PonceDeLeon:1988rg}
  J.~Ponce De Leon,
  Gen.\ Rel.\ Grav.\  {\bf 20}, 539 (1988).



\bibitem{Overduin:1998pn}
  J.~M.~Overduin and P.~S.~Wesson,
  Phys.\ Rept.\  {\bf 283}, 303 (1997)
  [arXiv:gr-qc/9805018].

\bibitem{Frolov:2003mc}
  V.~P.~Frolov, M.~Snajdr and D.~Stojkovic,
  Phys.\ Rev.\  D {\bf 68} (2003) 044002
  [arXiv:gr-qc/0304083].

\bibitem{Frolov:2004bq}
  V.~P.~Frolov, D.~V.~Fursaev and D.~Stojkovic,
  Class.\ Quant.\ Grav.\  {\bf 21}, 3483 (2004)
  [gr-qc/0403054].

\bibitem{Stojkovic:2004hp}
  D.~Stojkovic,
  Phys.\ Rev.\ Lett.\  {\bf 94}, 011603 (2005)
  [hep-ph/0409124].




\bibitem{add} N. Arkani-Hamed, S. Dimopoulos and G. R. Dvali,
Phys. Lett. B429, 263 (1998)



\bibitem{Starkman:2001xu}
  G.~D.~Starkman, D.~Stojkovic and M.~Trodden,
  Phys.\ Rev.\ Lett.\  {\bf 87}, 231303 (2001)
  [hep-th/0106143].

\bibitem{Starkman:2000dy}
  G.~D.~Starkman, D.~Stojkovic and M.~Trodden,
  Phys.\ Rev.\ D {\bf 63}, 103511 (2001)
  [hep-th/0012226].

\bibitem{rs} L. Randall and R. Sundrum,
 Phys. Rev. Lett. 83, 4690 (1999)




\bibitem{withKinney} D. Stojkovic and W. Kinney, work in progress




\bibitem{Martin:2000xs}
J.~Martin and R.~H.~Brandenberger,
``The trans-Planckian problem of inflationary cosmology,''
Phys.\ Rev.\ D {\bf 63}, 123501 (2001)
[arXiv:hep-th/0005209].

\bibitem{Niemeyer:2000eh}
J.~C.~Niemeyer,
``Inflation with a high frequency cutoff,''
Phys.\ Rev.\ D {\bf 63}, 123502 (2001)
[arXiv:astro-ph/0005533].


\bibitem{Brandenberger:2002hs}
R.~H.~Brandenberger and J.~Martin,
``On signatures of short distance physics in the cosmic microwave  background,''
Int.\ J.\ Mod.\ Phys.\ A {\bf 17}, 3663 (2002)
[arXiv:hep-th/0202142].

\bibitem{Burgess:2002ub}
C.~P.~Burgess, J.~M.~Cline, F.~Lemieux and R.~Holman,
``Are inflationary predictions sensitive to very high energy physics?,''
JHEP {\bf 0302}, 048 (2003)
[arXiv:hep-th/0210233].

\bibitem{Martin:2003kp}
  J.~Martin and R.~Brandenberger,
  ``On the dependence of the spectra of fluctuations in inflationary  cosmology
  on trans-Planckian physics,''
  Phys.\ Rev.\  D {\bf 68}, 063513 (2003)
  [arXiv:hep-th/0305161].

\bibitem{Schalm:2004qk}
K.~Schalm, G.~Shiu and J.~P.~van der Schaar,
``Decoupling in an expanding universe: Boundary RG-flow affects initial
conditions for inflation,''
JHEP {\bf 0404}, 076 (2004)
[arXiv:hep-th/0401164].


\bibitem{Easther:2001fi}
R.~Easther, B.~R.~Greene, W.~H.~Kinney and G.~Shiu,
``Inflation as a probe of short distance physics,''
Phys.\ Rev.\ D {\bf 64}, 103502 (2001)
[arXiv:hep-th/0104102].

\bibitem{Hui:2001ce}
L.~Hui and W.~H.~Kinney,
``Short distance physics and the consistency relation for scalar and  tensor fluctuations in the inflationary universe,''
Phys.\ Rev.\ D {\bf 65}, 103507 (2002)
[arXiv:astro-ph/0109107].

\bibitem{Easther:2001fz}
R.~Easther, B.~R.~Greene, W.~H.~Kinney and G.~Shiu,
``Imprints of short distance physics on inflationary cosmology,''
Phys.\ Rev.\ D {\bf 67}, 063508 (2003)
[arXiv:hep-th/0110226].

\bibitem{Easther:2002xe}
R.~Easther, B.~R.~Greene, W.~H.~Kinney and G.~Shiu,
``A generic estimate of trans-Planckian modifications to the primordial  power spectrum in inflation,''
Phys.\ Rev.\ D {\bf 66}, 023518 (2002)
[arXiv:hep-th/0204129].

\bibitem{Easther:2004vq}
  R.~Easther, W.~H.~Kinney and H.~Peiris,
  ``Observing trans-Planckian signatures in the cosmic microwave  background,''
  JCAP {\bf 0505}, 009 (2005)
  [arXiv:astro-ph/0412613].

\bibitem{Easther:2005yr}
  R.~Easther, W.~H.~Kinney and H.~Peiris,
  ``Boundary effective field theory and trans-Planckian perturbations:
  Astrophysical implications,''
  JCAP {\bf 0508}, 001 (2005)
  [arXiv:astro-ph/0505426].

\bibitem{Elgaroy:2003gq}
O.~Elgaroy and S.~Hannestad,
``Can Planck-scale physics be seen in the cosmic microwave background?,''
arXiv:astro-ph/0307011.

\bibitem{Okamoto:2003wk}
T.~Okamoto and E.~A.~Lim,
``Constraining Cut-off Physics in the Cosmic Microwave Background,''
Phys.\ Rev.\ D {\bf 69}, 083519 (2004)
[arXiv:astro-ph/0312284].

\bibitem{Jackson:2010cw}
  M.~G.~Jackson and K.~Schalm,
  ``Model Independent Signatures of New Physics in the Inflationary Power
  Spectrum,''
  arXiv:1007.0185 [hep-th].

\bibitem{Rinaldi:2010yp}
  M.~Rinaldi,
  Class.\ Quant.\ Grav.\  {\bf 29}, 085010 (2012)
  [arXiv:1011.0668 [astro-ph.CO]].

\bibitem{Stelle:1977ry}
  K.~S.~Stelle,
  Gen.\ Rel.\ Grav.\  {\bf 9}, 353 (1978).



\bibitem{Weinberg:1}  S.~Weinberg, in {\em Understanding the Fundamental Constituent of Matter} (ed. A. Zichichi, Plenum Press, New York, 1977).


\bibitem{Weinberg:2}
  S.~Weinberg,
in {\em General Relativity}, (eds. S. W. Hawking and W. Israel,  Cambridge University Press, 1979).


\bibitem{Gastmans:1977ad}
  R.~Gastmans, R.~Kallosh and C.~Truffin,
  Nucl.\ Phys.\  B {\bf 133}, 417 (1978);
  S.~M.~Christensen and M.~J.~Duff,
  Phys.\ Lett.\  B {\bf 79}, 213 (1978).

\bibitem{Reuter:1996cp}
  M.~Reuter,
  Phys.\ Rev.\  D {\bf 57}, 971 (1998)
  [arXiv:hep-th/9605030].

\bibitem{Codello:2006in}
  A.~Codello and R.~Percacci,
  Phys.\ Rev.\ Lett.\  {\bf 97}, 221301 (2006)
  [arXiv:hep-th/0607128];
  A.~Codello, R.~Percacci and C.~Rahmede,
  Int.\ J.\ Mod.\ Phys.\  A {\bf 23}, 143 (2008)
  [arXiv:0705.1769 [hep-th]];
  A.~Codello, R.~Percacci and C.~Rahmede,
  Annals Phys.\  {\bf 324}, 414 (2009)
  [arXiv:0805.2909 [hep-th]].


\bibitem{Lauscher:2001ya}
  O.~Lauscher and M.~Reuter,
  Phys.\ Rev.\  D {\bf 65}, 025013 (2002)
  [arXiv:hep-th/0108040];
  O.~Lauscher and M.~Reuter,
  Phys.\ Rev.\  D {\bf 66}, 025026 (2002)
  [arXiv:hep-th/0205062];



\bibitem{Cheng:1987ga}
  H.~Cheng and T.~T.~Wu,
{\it Expanding Protons: Scattering at High Energies}, (MIT Press, Cambridge, Massachusetts, 1987);
L. N. Lipatov, {\it Review in Perturbative QCD}, (Ed. A. H. Muller, World Scientific, Singapore, 1989),
and references therein.


\bibitem{Lipatov:1988ce}
  L.~N.~Lipatov,
  Nucl.\ Phys.\  B {\bf 309}, 379 (1988);
  L.~N.~Lipatov,
  Nucl.\ Phys.\  B {\bf 365}, 614 (1991).



\bibitem{Nachtmann:1991ua}
  O.~Nachtmann,
  Annals Phys.\  {\bf 209} (1991) 436.



\bibitem{Verlinde:1993te}
  H.~L.~Verlinde and E.~P.~Verlinde,
  arXiv:hep-th/9302104;
  I.~Y.~Arefeva,
  Phys.\ Lett.\  B {\bf 325}, 171 (1994)
  [arXiv:hep-th/9311115];
  M.~Li and C.~I.~Tan,
  Phys.\ Rev.\  D {\bf 50}, 1140 (1994)
  [arXiv:hep-th/9401134];
  R.~Kirschner, L.~N.~Lipatov and L.~Szymanowski,
  Nucl.\ Phys.\  B {\bf 425}, 579 (1994)
  [arXiv:hep-th/9402010];
  D.~Y.~Ivanov, R.~Kirschner, E.~M.~Levin, L.~N.~Lipatov, L.~Szymanowski and M.~Wusthoff,
  Phys.\ Rev.\  D {\bf 58}, 074010 (1998)
  [arXiv:hep-ph/9804443];
  J.~Bartels, V.~S.~Fadin and L.~N.~Lipatov,
  Nucl.\ Phys.\  B {\bf 698}, 255 (2004)
  [arXiv:hep-ph/0406193].

\bibitem{Ambjorn:1998xu}
  J.~Ambjorn and R.~Loll,
  Nucl.\ Phys.\  B {\bf 536}, 407 (1998)
  [arXiv:hep-th/9805108].


\bibitem{Hawking:1978jz}
  S.~W.~Hawking,
  Phys.\ Rev.\  D {\bf 18}, 1747 (1978).


\bibitem{David:1984tx}
  F.~David,
  Nucl.\ Phys.\  B {\bf 257}, 45 (1985);
  Nucl.\ Phys.\  B {\bf 257}, 543 (1985);
  J.~Ambjorn, B.~Durhuus and J.~Frohlich,
  Nucl.\ Phys.\  B {\bf 257}, 433 (1985).





\bibitem{Ambjorn:2000dv}
  J.~Ambjorn, J.~Jurkiewicz and R.~Loll,
  Phys.\ Rev.\ Lett.\  {\bf 85}, 924 (2000)
  [arXiv:hep-th/0002050].


\bibitem{Ambjorn:2005db}
  J.~Ambjorn, J.~Jurkiewicz and R.~Loll,
  Phys.\ Rev.\ Lett.\  {\bf 95}, 171301 (2005)
  [arXiv:hep-th/0505113].

\bibitem{Lauscher:2005qz}
  O.~Lauscher and M.~Reuter,
  JHEP {\bf 0510}, 050 (2005)
  [arXiv:hep-th/0508202].



\bibitem{horava} P.~Horava, ``Quantum Gravity at a Lifshitz Point,'' Phys.~Rev.~{\bf D 79}, 084008 (2009)
\bibitem{massive} W.~de Paula, O.~Miranda, R.~Marinho, ``Polarization states of gravitational waves with a massive graviton,'' Class.~Quant.~Grav. {\bf 21}, 4595 (2004)

\bibitem{withNiayesh}
  N.~Afshordi and D.~Stojkovic,
  arXiv:1405.3297 [hep-th].

\bibitem{Hao:2014tsa}
  P.~Hao and D.~Stojkovic,
  arXiv:1404.7145 [gr-qc].



\bibitem{Carlip:2009km}
  S.~Carlip,
  arXiv:1009.1136 [gr-qc].
\bibitem{Carlip:2011tt}
  S.~Carlip, R.~A.~Mosna and J.~P.~M.~Pitelli,
  Phys.\ Rev.\ Lett.\  {\bf 107}, 021303 (2011)
  [arXiv:1103.5993 [gr-qc]].



\bibitem{lmpn1} L.~Modesto and P.~Nicolini, Phys.~Rev.~D~{\bf 81}, 104040 (2010) [arxiv:0912.0220 [hep-th]].
%
\bibitem{Shirkov:2010sh}
  D.~V.~Shirkov,
  Part.\ Nucl.\ Lett.\  {\bf 7}, 625 (2010)
  [arXiv:1004.1510 [hep-th]].




\bibitem{Ivanenko:1992qw}
  I.~P.~Ivanenko, V.~V.~Kopenkin, A.~K.~Managadze and I.~V.~Rakobolskaya,
  JETP Lett.\  {\bf 56}, 188 (1992);
  V.~V.~Kopenkin, A.~K.~Managadze, I.~V.~Rakobolskaya and T.~M.~Roganova,
  Phys.\ Rev.\  D {\bf 52}, 2766 (1995)
  [arXiv:hep-ph/9408247].


\bibitem{Xue:1999bb}
  L.~Xue {\it et al.},
in {\it Proceedings of the 26th International Cosmic Ray Conference},
Salt Lake City, {\bf 1} 127 (1999)

\bibitem{Capdevielle:1988pe}
  J.~N.~Capdevielle,
  J.\ Phys.\ G {\bf 14}, 503 (1988);
  A.~K.~Managadze,
  Part.\ Nucl.\ Lett.\  {\bf 112}, 19 (2002);
  V.~I.~Galkin,
  Bull.\ Russ.\ Acad.\ Sci.\ Phys.\  {\bf 66}, 1697 (2002)
  [Izv.\ Ross.\ Akad.\ Nauk.\  {\bf 66} (2002) 1544];
A.~K. Managadze and V. I. Osedlo,
Bull.\ Russ.\ Acad.\ Sci.\ Phys.\  {\bf 73}, 615 (2009)
  [Izv.\ Ross.\ Akad.\ Nauk Ser.\ Fiz.\  {\bf 73}, 653 (2009)].




\bibitem{Apanasenko:icrc}
  A.~V.~Apanasenko {\it et al.}
in
{\it Proceedings of the 15th International Cosmic Ray Conference},
Plovdiv, {\bf 7} 220 (1977);
A.~K.~Managadze {\it et al.},
in
{\it Proceedings of the 27th International Cosmic Ray Conference},
Hamburg, {\bf 1} 1426 (2001);
A.~K. Managadze {\it et al.},
Bull.\ Russ.\ Acad.\ Sci.\ Phys.\  {\bf 71N4}, 513 (2007)
  [Izv.\ Ross.\ Akad.\ Nauk Ser.\ Fiz.\  {\bf 71}, 530 (2007)];
A.~K. Managadze and V. I. Osedlo,
Bull.\ Russ.\ Acad.\ Sci.\ Phys.\  {\bf 73N5}, 617 (2009)
  [Izv.\ Ross.\ Akad.\ Nauk Ser.\ Fiz.\  {\bf 73}, 653 (2009)].

\bibitem{Deile:2010mv}
A. De Roeck, I. P. Lokhtin, A. K. Managadze, L. I. Sarycheva, and
A. M. Snigirev,
in {\it Proceedings of the 13th International Conference on Elastic
  and Diffractive Scattering} (Edts. M. Deile, D. d'Enterria, and
A. De Roeck, 2010) p.308;
M.~Deile {\it et al.},
  arXiv:1002.3527 [hep-ph];
R.~Mukhamedshin,
Eur. Phys. J. C {\bf 60}, 345 (2009). See also Ref.~\cite{Ivanenko:1992qw}.



\bibitem{Calcagni:2009kc}
  G.~Calcagni,
  Phys.\ Rev.\ Lett.\  {\bf 104}, 251301 (2010);

\bibitem{Calcagni:2010bj}
  G.~Calcagni,
  JHEP {\bf 1003}, 120 (2010)

\bibitem{Nicolini:2011nz}
  P.~Nicolini and E.~Winstanley,
  JHEP {\bf 1111}, 075 (2011)
  [arXiv:1108.4419 [hep-ph]].

\bibitem{Calcagni:2011sz}
  G.~Calcagni,
  JHEP {\bf 1201}, 065 (2012)
  [arXiv:1107.5041 [hep-th]].


\bibitem{Modesto:2011kw}
  L.~Modesto,
  Phys.\ Rev.\ D {\bf 86}, 044005 (2012)
  [arXiv:1107.2403 [hep-th]].

\bibitem{Obukhov:2011ks}
  Y.~N.~Obukhov, A.~J.~Silenko and O.~V.~Teryaev,
  Phys.\ Rev.\ D {\bf 84}, 024025 (2011)
  [arXiv:1106.0173 [hep-th]].

\bibitem{Mann:2011rh}
  R.~B.~Mann and J.~R.~Mureika,
  Phys.\ Lett.\ B {\bf 703}, 167 (2011)
  [arXiv:1105.5925 [hep-th]].

\bibitem{Nieves:2011fy}
  J.~F.~Nieves,
  Int.\ J.\ Mod.\ Phys.\ A {\bf 26}, 5387 (2011)
  [arXiv:1105.2546 [hep-ph]].

\bibitem{Mureika:2011py}
  J.~R.~Mureika and P.~Nicolini,
  Phys.\ Rev.\ D {\bf 84}, 044020 (2011)
  [arXiv:1104.4120 [gr-qc]].

\bibitem{Landsberg:2010zz}
  G.~Landsberg,
  PoS ICHEP {\bf 2010}, 399 (2010).

\bibitem{Stoica:2013wx}
  O.~C.~Stoica,
  arXiv:1301.2231 [gr-qc].

\bibitem{GonzalezMestres:2010pi}
  L.~Gonzalez-Mestres,
  arXiv:1009.1853 [astro-ph.HE].

\bibitem{Calmet:2010vp}
  X.~Calmet and G.~Landsberg,
  chapter 7 in A.J. Bauer and D.G.Eiffel editors,Black Holes: Evolution, Theory and Thermodynamics Nova Publishers, New York, 2012
  [arXiv:1008.3390 [hep-ph]].


\bibitem{Caravelli:2010be}
  F.~Caravelli and L.~Modesto,
  Phys.\ Lett.\ B {\bf 702}, 307 (2011)
  [arXiv:1001.4364 [gr-qc]].




\end{thebibliography}
\end{document}